\documentclass[portrait, 6pt]{article}
\usepackage[utf8]{inputenc}
\usepackage{changepage}
\usepackage[margin=1in]{geometry}
\usepackage[font= Large,labelfont=bf]{caption}
\usepackage{setspace}
\usepackage{relsize}
\usepackage{amsmath}
\usepackage{rotating}
\usepackage[english]{babel}
\usepackage[autostyle, english = american]{csquotes}
\MakeOuterQuote{"}

\usepackage{etoolbox} 
\makeatletter 
\patchcmd{\ps@headings}{{\slshape\rightmark}\hfil\thepage}{\thepage\hfil}{}{}
\makeatother
\usepackage{verbatim}

\usepackage{booktabs}
\usepackage{longtable}
\usepackage{array}
\usepackage{multirow}
\usepackage{wrapfig}
\usepackage{float}
\usepackage{bbm}
\usepackage{colortbl}
\usepackage{lscape}
\usepackage{pdfpages}
\usepackage{lscape}
\usepackage{tabu}
\usepackage{threeparttable}
\usepackage{threeparttablex}
\usepackage[normalem]{ulem}
\usepackage{makecell}
\usepackage{xcolor}

\usepackage{natbib}
\usepackage{url}

\renewenvironment{abstract}
 {\small
  \begin{center}
  \bfseries \abstractname\vspace{-.5em}\vspace{0pt}
  \end{center}
  \list{}{
    \setlength{\leftmargin}{1cm}%
    \setlength{\rightmargin}{\leftmargin}%
  }%
  \item\relax}
 {\endlist}

\begin{document}

\title{\textbf{Powering Up a Slow Charging Market: How Do Government Subsidies Affect Charging Station Supply?}}
\author{Zunian Luo\thanks{Weston High School, Weston, MA, 02493; MIT JTL Urban Mobility Lab, 77 Massachusetts Ave. MIT 9-523, Cambridge, MA, 02139. E-mail: 23luoz@weston.org. Special thanks to my mentors Yunhan Zheng (MIT) and Jinhua Zhao (MIT) at the JTL Urban Mobility Lab and Bhargav Gopal (Columbia) for their invaluable feedback and guidance throughout the research process. I am grateful to Robert Barro (Harvard), Christopher Knittel (MIT Sloan), James Poterba (MIT), David Laibson (Harvard), Shanjun Li (Cornell Dyson), Michael Greenstone (UChicago), Robert Stavins (Harvard Kennedy), Lucas Davis (Berkeley Haas), Wolfram Schlenker (Columbia SIPA), Eileen Appelbaum (CEPR), and an anonymous referee (MIT) for their insightful feedback and suggestions of outlets for scholarly dissemination.
}}
\date{\today}

\maketitle

\begin{abstract}Electric vehicle adoption is considered to be a promising pathway for addressing climate change. However, the market for charging stations suffers from a market failure: a lack of EV sales disincentives charging station production, which in turn inhibits mass EV adoption. Charging station subsidies are often discussed as policy levers that can stimulate charging station supply and correct this market failure. Nonetheless, there is limited research examining the extent such subsidies are successful in promoting charging station supply. Using annual data on electric vehicle sales, charging station counts, and subsidy amounts from 57 California counties and a staggered difference-in-differences methodology, I find that charging station subsidies are highly effective: counties that adopt subsidies experience a 36\% increase in charging station supply 2 years following subsidy adoption. This finding suggests that governmental intervention can help correct the market failure in the charging station market. 

\end{abstract}

\vspace{1em}
\noindent
Keywords: Electric Vehicles, Difference in Differences, Climate Change \\
JEL classification: Q48, Q58, L62
\\ 
\\
\\
\newpage

\section{Introduction}
Amid concern over the heightening social and economic costs of climate change, policymakers and the public alike are demanding cost-effective climate solutions.\footnote{The US, for example, has already suffered more than two trillion dollars in climate change related damages (NCEI, 2022).}  The electrification of the light-duty vehicle (LDV) fleet offers one promising pathway to reducing emissions in the transportation sector -- the source of around 29\% of the US’ carbon emissions (World Resources Institute, 2019).\footnote{Different from ICEs, which consume petroleum, EVs may indirectly produce emissions from electricity stored in rechargeable batteries and when operated in all-electric mode, produce zero tailpipe greenhouse gas (GHG) emissions. In addition to emitting 60\% fewer GHG emissions, EVs emit, on average, an estimated 61\% less volatile organic compounds, 93\% less carbon monoxide, 28\% less nitrogen oxides, and 32\% less black carbon than ICEs (Lattanzio and Clark, 2020; Hall and Lutsey, 2018);  Holland et al. (2016) argue that EV subsidies should be negative in some parts of the US to account for regional variations in sources of electricity generation. An EV in West Virginia, for example, can indirectly produce more carbon emissions than ICEs due to the region’s high reliance on coal as a source of electricity. Nonetheless, the net carbon footprint of electricity is steadily declining thanks to the growing shift to clean sources of electricity such as natural gas, solar, and nuclear power (EIA, 2021).}  Recognizing this potential, federal and state policymakers have set ambitious long-term goals to replace internal combustion engines (ICEs) with electric vehicles (EVs), in conjunction with programs meant to stimulate charging station supply.\footnote{In 2021, President Biden issued an executive order targeting 50\% nationwide EV sales share in 2030 (The White House, 2021). California leads the way on the state level, requiring that, by 2035, all new cars and passenger trucks sold in California be zero-emission vehicles (Office of Governor Newsom, 2020). In regard to charging station programs, the 2021 Infrastructure Investment and Jobs Act dedicates \$7.5 billion to the construction of a comprehensive nationwide network of public EV chargers, and the Alternative Fuel Infrastructure Tax Credit (Infrastructure and Investment and Jobs Act, 2021; The White House, 2021). The Alternative Fuel Infrastructure Tax Credit provides a tax credit of 30\% of the total cost, not to exceed \$30,000, for fueling equipment for natural gas, propane, liquefied hydrogen, electricity, E85, or diesel fuel blends containing a minimum of 20\% biodiesel installed through December 31, 2021. Presently, the program has expired (AFDC, 2022).}  But even as the importance of charging stations in promoting EV adoption comes to the fore, research on how to stimulate charging station production remains limited.

Economic theory indicates that producer subsidies, which lower the out-of-pocket cost of constructing new charging stations, should increase the supply of stations, especially when supply is elastic. But with EV share of the total vehicle market at 5\%, private actors are inadequately incentivized to construct charging stations.\footnote{See Table 2.} Range anxiety due to the lack of charging station infrastructure leads consumers not to purchase EVs.\footnote{As a relatively recent technology, EVs contend with several obstacles to widespread adoption, including an inaccessible purchase price, limited range, and long charging times (Bakker and Trip, 2013; Egbue and Long, 2012; She et al., 2017; Zhang et al 2018). The latter two factors can inspire range anxiety, which in turn may lead to reluctance to purchase EVs among prospective consumers, particularly in “charging deserts'' where public stations are scarce (Chen, Kockelman, and Hanna, 2019; Desai et al., 2021; Hardman et al. 2021; Meunier and Ponssard, 2020; Greene et al., 2020).}  In turn, charging station providers facing limited EV demand may find it unprofitable to invest in a large charging station network. This interdependence between charging stations and EV adoption generates a market equilibrium where the quantity of charging stations and EVs is less than socially optimal. Figure 1 highlights this interdependence; DC Stations and EV sales trend relatively similarly throughout the observed period. Growth in DC stations lags growth in EV sales, consistent with the theory that station providers consider market conditions when deciding to build charging stations; it is possible, for example, that station producers take time to respond to growing EV sales by constructing new charging stations.\footnote{Agents in counties that have greater EV sales may also have a more positive social attitude towards climate change action and clean transportation in particular.} In the most extreme case, the EV market can unravel as a lack of charging stations dissuades consumers from purchasing EVs, which further disincentives suppliers from producing charging stations. This theoretical motivation, together with the lack of empirical evidence on the effects of charging station subsidies, provide impetus for the following question: to what extent do government charging station subsidies increase charging station supply?\footnote{This question has become of great concern particularly in recent years as federal and state governments ramp up deployment of charging station subsidies (President Biden, USDOT and USDOE Announce \$5 Billion Over Five Years for National EV Charging Network, Made Possible by Bipartisan Infrastructure Law, n.d.).}

In this paper, I employ a staggered difference-in-differences approach to evaluate how charging station subsidies affect charging station supply at the county level. Using a rich data set of charging station counts obtained from the National Renewable Energy Laboratory and manually compiled data sets of purchase price and charging station subsidy counts, I quantify the elasticity of charging station supply with respect to charging station subsidies. I leverage the fact that 26 California counties adopted charging station subsidies between 2010 and 2022. I then compare changes in charging station supply, EV sales measures, and county demographic information in these “treated” counties to changes in “untreated” counties — those that did not adopt charging station subsidies between 2010 and 2022.\footnote{The total sample of 57 California counties is split between 26 "treated" counties and 31 "untreated" counties} 

The primary motivation for the analysis is to determine if charging station subsidies are successful at expanding charging station supply, and, if so, where they are most effective. If charging station supply is highly elastic with respect to charging station subsidies, policymakers may consider increasing public funds for such subsidies worthwhile. Understanding this elasticity can also enable policymakers to forecast the amount of subsidies needed to meet a particular charging station deployment objective. 

To identify the causal effects of the subsidy, I examine how county-level outcomes of all treated counties evolve in the treatment period from 2010 to 2022. Comparing treated counties to untreated counties, I verify that the conditional independence assumption holds.\footnote{The assertion that the conditional independence holds is supported by the fact that $\hat{\delta^{\tau}}$, $-12 \le \tau \le -3 \approx 0$ for charging station supply.} I employ an event study specification, where the subsidy's effect varies by year, and a staggered difference in difference model where the effect is assumed to be immediate and constant. Since $\hat{\beta}^t$ may be contaminated by heterogeneous cohort-specific treatment effects, I reproduce the event study with an alternative interaction weighted estimator proposed by Sun and Abraham (2021) as a robustness check. While I verify the conditional independence assumption is likely to hold for charging stations in the Sun and Abraham estimation, it is not likely to hold for EV sales; sales tend to increase in treated counties even before subsidy adoption. Thus, I assign causal interpretation on the impact of subsidies on charging station supply but not on EV sales.

Across various specifications, the staggered difference in differences analysis finds statistically significant effects of charging station subsidies on charging station supply: a 100\% increase in charging station subsidies increases charging station supply by 2.5\%. Elasticity interpretations from the staggered difference in differences model, otherwise known as a constant elasticity model, are independent of initial subsidy value. However, the constant elasticity model may be skewed in the presence of many zeros (Boulton and Williford, 2018; Green, 2021). Given this consideration, I study the impact of subsidy adoption using the model proposed by Sun and Abraham (2021). 

This paper contributes to several literatures. First, the literature on charging stations has focused mainly on the direct causal effect of charging stations and charging station subsidies on EV sales (Springel, 2021; Li et al., 2017; Remmy, 2022; van Dijk, Delacrétaz, and Lanz, 2022; Yu, Li, and Tong, 2016; Fournel, 2022; Cole et al., 2021; Ma and Fan, 2020; Ledna et al., 2022; Gnann, Plötz, and Wietschel, 2019). My study offers a first analysis to quantify the elasticity of charging station supply with respect to charging station subsidies. More broadly, I contribute to the literature on the role of transportation sector government policy in mitigating climate change (for an overview, see Rapson and Muehlegger, 2021).\footnote{Within the literature on clean energy vehicle diffusion, empirical studies primarily focus on the effect of purchase price subsidies on EV sales (Gallagher and Muehlegger, 2011; Gong, Ardeshiri, and Rashidi, 2020; Li et al., 2019; Diamond, 2009; Hardman et al., 2017).}  Graff Zivin et al (2014), Archsmith et al. (2015), and Holland et al. (2016) find that EVs can emit substantially less lifetime carbon emissions and local pollutants than ICEs (depending on regional electricity sources). On a geopolitical level, Ajanovic and Haas (2018) and Fernández et al. (2011) find that replacing ICEs with EVs has important benefits for energy security by reducing reliance on gasoline imports. Increasing charging station supply for EV consumers can be an important factor in realizing these external benefits.

Second, I contribute to the emerging literature on socioeconomic dimensions of charging station supply by exploring heterogeneity in the effect of charging station subsidies across various socioeconomic dimensions. Hsu and Fingerman (2021) document disparities in public charging station access across income and race, motivating study of how the effects of charging station subsidies vary among different treatment groups. Disadvantaged communities persistently have less access to charging compared to their wealthier counterparts because of their lower capacity to afford EVs and inability to utilize home charging solutions (Nicholas, Hall, and Lutsey, 2019; Pierce, McOmber, and DeShazo, 2020; Axsen and Kurani, 2012; Lopez-Behar et al, 2019; Khan et al., 2022).\footnote{Charging at home is a popular alternative to public chargers because it is the cheapest and most convenient method of charging. But such an option is typically only accessible for wealthier residents who often live in single or multi-family homes that are spatially conducive to such residential chargers. Low-income residents, meanwhile, rely more on public charging because they tend to live in large multi-unit dwellings (MUDs) which often lack adequate residential charging (Wei et al, 2021; Muehlegger and Rapson, 2018). Installing residential chargers in rental residences poses a greater challenge as renters have less incentive to bear the cost of upgrading a home they themselves do not own and owners have less incentive to bear the cost of a charger they may not use.} For these regions, ensuring adequate public charging access is especially crucial (Dunckley and Tal, 2016; Funke et al., 2019; Axsen and Kurani, 2012; C2ES, 2017; Canepa, Hardman, and Tal, 2019).\footnote{Even while rising nationwide EV sales and charging station supply may appear to signal promising trends, they mask the diverging paths of EV ownership and charging stations access along socioeconomic dimensions; wealthy whites who can easily access charging are the main impetus behind accelerating EV ownership, while low income Blacks and Hispanics living in charging deserts are left in the dust.} 

Third, recent studies document indirect network effects $-$ there is interdependence between charging station supply and EV adoption (Springel, 2021; Li et al., 2017). Accounting for these indirect network effects, studies estimating the causal relationship between charging stations and EV sales use an instrumental variable methodology (Springel, 2021; Remmy, 2022; Li et al., 2017; van Dijk, Delacrétaz, and Lanz, 2022; Fournel, 2022; Pavan, 2017). A similar relationship could exist between charging station subsidies and EV sales. While several studies have examined endogeneity in the relationship between charging infrastructure supply and EV adoption, there is scant investigation on how sales measures evolve prior to the adoption of charging station subsidies (Egnér and Trosvik, 2018; Sierzchula et al., 2014). I find that counties that adopt charging station subsidies tend to have faster EV sales growth trajectories than non-adopting counties prior to subsidy adoption. Uncovering this endogeneity is important because it could bias typical difference-in-differences estimates; if one incorrectly assumed the parallel trends assumption holds, estimates would be overestimated by 11.5\%.\footnote{To obtain this value, I find the absolute value of the average of the estimate at the binned value from $\tau$ = -4 to $\tau$ = -12 and the estimate at $\tau = -2$ (-0.21 and -0.02, respectively) in Table 6} Further, understanding the presence of significant pre trends suggests future studies that examine the causal relationship between charging station subsidies and EV sales should utilize an instrumental variable design. 

The paper proceeds as follows. Section 2 describes the background on charging stations and subsidy policy. Section 3 details data, descriptive statistics, and bivariate relationships. Section 4 presents the econometric specifications. Section 5 discusses the results and robustness checks. Section 6 concludes.

\section{Institutional Context}
\subsection{Industry Background}
Consumers may consider adequate access to charging stations an important factor in deciding whether to purchase an EV. Charging stations are split into three levels — Level 1, Level 2, and DC — and may be employed either in a residential or public setting. Levels 1 and 2 charging use a universal connector that can be plugged into any EV.\footnote{Tesla is an exception in that Tesla Level 2 stations, like their DC stations, use a proprietary Tesla connector. However, the recent increase in available aftermarket adapters have made Tesla chargers more compatible with other car models.} DC stations, meanwhile, have three types of plugs: CHAdeMO, SAE Combo (CCS), and Tesla.\footnote{The CHAdeMO plug is compatible with Nissan, Mitsubishi, Kia, Fuji, and Toyota brand models; SAE Combo (CSS) is compatible with U.S. and European EVs, BMW, Volkswagon, Chevy, and some Asian electric vehicles; Tesla plugs are compatible with Tesla vehicles. See https://blinkcharging.com/understanding-ev-charging-plugs/?locale=en} Although DC stations are not homogeneous, the growing market of charging plug adapters is increasing charging flexibility among EV owners. 

Charging speed is the main difference between these three types of stations. DC chargers are the fastest type of charging station, outputting typically between 200 – 600 V, and can fully charge an EV in less than an hour (California Energy Commission, n.d). In addition to their fast charging, DC stations are almost exclusively employed in either public or private commercial settings due to the significant cost of investment required.\footnote{According to a study from the International Council on Clean Transportation, DC chargers typically cost from \$28,000 to \$140,000 to install, excluding long term maintenance costs (Nicholas, 2019).} Level 2 chargers are faster, outputting 240 V, and can fully charge an EV overnight; they are the most ubiquitous type of charging station and are commonly used in both commercial and residential settings. Due to their superior charging speed and importance in alleviating range anxiety, I focus on the effect of DC subsidies on DC station supply. Results and discussion on the effect of Level 2 charging subsidies on Level 2 stations can be found in the Appendix.\footnote{Level 1 is the slowest, which outputs 120 V of electricity, are almost exclusively used by households, and are considered “trickle chargers” because it takes around 24 hours to fully charge an empty battery. Level 1 chargers are the least costly of the three types because they are often included in the purchase of an EV and can plug directly into a standard 120 V outlet. I exclude Level 1 chargers from the analysis because the population of public Level 1 chargers is extremely limited.} Although the majority of charging station investors are private firms specializing in charging station networks, such as ChargePoint and Blink, governments from city and county to the federal level, as well as utilities corporations, may also directly invest in charging station construction.\footnote{San Diego Gas and Electric, and Pacific Gas and Electric, for instance, have programs that directly invest in charging stations (Armenia \& Johnson, 2022; EV Charge Program, (n.d.)).}

There are two major types of EVs: battery electric vehicles (BEVs) which are fully electric and derive their power entirely from batteries (e.g.Tesla Model 3), and plug-in hybrid vehicles (PHEVs) which obtain power from batteries and another fuel source, typically gasoline (e.g. Toyota Prius). I group EV sales data by PHEV sales, BEV sales, and EV sales — an aggregate measure of PHEV and BEV sales. Although growing EV sales may reflect an expanding EV fleet, it does not necessarily indicate that consumers are replacing ICEs with EVs.\footnote{For example, a family that owns an ICE may purchase another EV to add to the household fleet. While the new EV appears as an additional “EV sale,” it does not necessarily take an ICE off the road — the hypothetical family could continue to drive their existing ICE. Since the average EV owner has a relatively high income, I expect that there are a substantial number of these cases where EV purchases do not actually serve to fully replace an ICE.} To address this nuance while examining cross sectional differences between treated and untreated counties (see Table 2), I supplement EV sales with ICE and EV share data obtained from the California Energy Commission since vehicles included in ICE and EV share must be “on the road.” 

\subsection{Subsidy Policy}
Charging station subsidies are among the most recent EV incentives to gain prominence. There are three different types of charging station incentives, each targeting a particular charging station type (i.e Level 1, Level 2, DC). Charging station subsidies are predominantly implemented at the state and county level rather than the federal level. Eligibility requirements vary by state and county, but most California counties require applicants to be registered as a California business or eligible organization. While few subsidy programs restrict access to manufacturers with a charging network size below a certain threshold, programs typically have limits on the amount of incentives available per station, which varies widely by station type. DC subsidies, for instance, vary around \$25,000 to \$80,000.\footnote{Level 2 subsidies may range from around \$500 to \$7,500.} Some subsidy programs, such as the statewide California Electric Vehicle Infrastructure Project (CALeVIP), offer increased rebate amounts for charging stations built in disadvantaged or low income communities (see data dictionary for more details on disadvantaged and low income classification). Subsidies reimburse businesses through a tax rebate or credit after the point of sale.

Table 1 displays the evolution of counties that adopt subsidies between 2010 and 2022. Very few counties adopted charging station subsidies between 2010 and 2016. The number of counties that adopt charging subsidies increased substantially from 2016-2021. Beginning in 2016, the trend in DC stations closely mirrors that of Level 2 subsidies, suggesting that counties tend to implement Level 2 and DC stations concurrently.\footnote{For the 2021 cohort, I do not identify event study results in relative years greater than one because they extend beyond the observation period. This limitation further motivates the staggered difference-in-differences estimation, which can be used to extract effects for late adopters. Original data sources on charging station and charging station subsidy counts can be found in the legal appendix.}

\section{Data and Descriptive Statistics}
\subsection{Data Description}
I link three data sets for California counties organized at the county and year level. (1) data on charging station counts by type (i.e Level 2, DC) is obtained from the National Renewable Energy Laboratory developer interface (National Renewable Energy Laboratory, 2022). The data set includes the station opening date and charging network status, allowing me to distinguish between Tesla and non Tesla stations. (2) EV sales data is obtained from the California Energy Commission (California Energy Commission, 2022). Within the county year level, sales are organized by make, model, and fuel type, enabling me to separate sales into PHEV and BEV categories. (3) I manually conduct a legal analysis of charging station subsidy amounts by type. I obtain subsidy amounts for the legal analysis from a variety of subsidy program web pages (see legal appendix). In addition to charging station subsidies, I compile county-year level data on purchase price subsidies to be used as a control variable in later empirical analyses.\footnote{Although some subsidies may have changed in magnitude prior to their current incentive amount, very few provide information on the size of subsidy magnitude changes and when they occurred, rendering it impossible to incorporate all subsidy magnitude changes on a per subsidy basis in the legal analysis.} I link observations at the county and year level, permitting me to study the relationship between charging station subsidies and charging stations, and between charging station subsidies and EV sales.

In all three datasets, I restrict attention to the sample period starting in 2010 and ending in 2022. Under this sample restriction, I observe the county profile, charging station counts, charging subsidy amounts, and EV sales measures of 57 California counties, which represents 98.3\% of the universe of all California counties. I expand data on charging station subsidies by county-year slightly by contacting subsidy program staff in counties where the necessary information is not publicly available. 

The legal analysis and charging station counts data are divided into two primary groups — Level 2 and DC — for which I assess subsidy effects separately. The DC group includes the number of DC stations and DC station subsidies in a given county in a given year.\footnote{Analogously, the Level 2 group includes the supply of Level 2 stations and Level 2 subsidies in a given county in a given year. Results of the analysis with Level 2 stations can be found in the appendix.} I explore heterogeneity across various dimensions. Economic theory suggests that the subsidy should have a greater effect on areas with high population density and high aggregate GDP; subsidies may have a less significant effect in regions with low population density because charging station producers may find it unprofitable to construct stations if there is a limited demand nearby. Equity considerations motivate investigating the subsidy's impact in disadvantaged communities.\footnote{Notes on Table 4 describe the criteria for each designation in detail.} To enable event study analysis, I group counties according to a dummy variable indicating treatment status: treated counties adopt charging station subsidies within the period from 2010 - 2022 while untreated counties do not adopt any such subsidies in the same period.\footnote{I assess differences in responsiveness to the subsidy by further separating between aggregate Level 2 stations and Tesla Destination Stations, and between aggregate DC stations and Tesla Superchargers. Tesla operates a substantial of number of Destination (Level 2) and Supercharger (DC) stations, but faces significantly different incentives compared to other private investors: since the business models of dedicated charging station suppliers such as Chargepoint, Blink, and Electrify America revolve around collecting fees from operating charging stations, I expect that they are relatively more sensitive to changes in the cost of constructing charging stations. Meanwhile, since Tesla vehicles can only charge at Tesla’s network of proprietary stations at present, Tesla has a significant incentive to ensure adequate charging access for areas of high demand, whether subsidies exist in that region. I am the first in the literature to examine how the effect of the subsidy differs between Tesla and non-Tesla stations.}

A unique strength of this paper is that it utilizes county level data from California, the leading state in EV adoption and charging infrastructure in the US.\footnote{Approximately 39\% of EVs nationwide are registered in the state of California. See afdc.energy.gov/data/10962} There are three pertinent advantages of using California as the observational setting: (1) California is leading state in terms of the amount of charging subsidies and stations, making it a fertile research ground to study the role of subsidies in promoting development of stations; (2) California has the largest EV market in the US, suggesting that the results provide guidance on the effect of subsidies in the event other states adopt similar subsidies down the road; (3) many counties within California do not adopt subsidies, and these counties plausibly provide a valid counterfactual.

I collect data on county-level population from the California Department of Finance. From the US Bureau of Economic Analysis, I retrieve GDP figures for each county from 2010 - 2020. Rural and disadvantaged county designations are obtained from the California State Association of Counties and the California Office of Environmental Health Hazard Assessment, respectively.

\subsection{Descriptive Statistics}
Treated counties are defined to be counties that adopt charging station subsidies between 2010 and 2022. Consistent with this definition, Column 3 in Table 2 shows that treated counties, on average, implement subsidies valued at \$43,044 while untreated counties do not receive any subsidies over the sample period. As seen in Column 3, there are also cross sectional differences in charging stations, sales measures, market characteristics, and demographic characteristics between treated and untreated counties. Interestingly, the fact that treated counties have substantially more purchase price subsidies than untreated counties suggests that counties that implement charging station subsidies tend to enact purchase price subsidies as well, perhaps as part of a broader favorable view of government support for EVs. Nonetheless, these cross sectional differences do not pose a problem for my identification strategy, which relies on the assumption that treatment and control counties trend similarly before subsidy adoption, not necessarily that treatment and control counties have similar profiles on average (i.e, parallel trends).

\subsection{Bivariate Relationships}
Figure 2 Panel A displays a positive, statistically significant correlation between charging station subsidies and charging station supply. Panel B, which has a larger estimate (0.23), likewise shows a positive, statistically significant correlation between charging station subsidies and EV sales. Notably, Panel A and Panel B have very similar distributions, demonstrating that counties with more charging station subsidies also tend to have a greater amount of charging stations. While the clear and significant correlations in Panels A and B do not necessarily indicate causality, they do motivate further investigation exploring whether the relationships depicted in Figure 2 are causal. 
 
Figure 3 identifies potential sources of omitted variable bias (OVB): to be considered a source of omitted variable bias, the variable in question must be extraneous to the model and correlated with the independent variable and dependent variable. Recall the two primary relationships discussed in this paper are those between charging station subsidies and charging stations and between charging station subsidies and EV sales. Thus, Figure 3 plots two potential sources of OVB — purchase price subsidies and GDP — against the three primary independent variables of interest in my analysis: charging station subsidies, charging stations, and EV sales. 

The first row of Figure 3 (Panels A, B, and C) displays a positive, statistically significant relationship between GDP and each of the three independent variables, demonstrating that GDP is correlated with charging station subsidies, charging stations, and EV sales. Wealthier consumers have a greater ability to pay for EVs, and as the EV fleet size increases in turn, so too should the private incentive to construct additional charging stations. Meanwhile the second row (Panels D, E, and F) display a positive, statistically significant, albeit more scattered relationship between purchase price subsidies and the three independent variables. While purchase price subsidies are strongly correlated with charging station subsidies, stations, and sales, I expect they have a greater influence on sales compared to stations because they directly reduce the price of purchasing a new EV, directly resulting in greater EV sales. On the other hand, purchase price subsidies do not lower the out of pocket costs of constructing a new charging station.\footnote{It should be noted that purchase price subsidies may indirectly affect charging station supply by stimulating EV sales.} To address these concerns, I control for GDP and purchase price subsidies in later OLS estimates.


\section{Empirical Strategy}

My empirical framework aims to accomplish the following: (1) I provide initial estimates from a standard staggered differences-in-differences regression; (2) I conduct an event study robustness check to verify the parallel trends assumption; and (3) I implement an interaction weighted estimator proposed by Sun and Abraham (2021) to determine if treatment effects are robust to heterogeneous cohort specific effects, where cohorts are defined by calendar year (e.g 2014). Notably, the independent variable in the Sun and Abraham estimation is binary.  

\subsection{Staggered difference-in-differences}
In my empirical framework, I estimate a generalized staggered difference-in-differences equation that can be applied to examine the effect of charging station subsidies on charging infrastructure supply and EV sales (Beck et al., 2010; Hoynes and Schanzenbach, 2009; Baker, Larcker, and Wang, 2022; Hoynes et al., 2016; Fauver et al., 2017; Wang et al., 2021):

\setlength{\parindent}{20pt}

\begin{equation}\label{staggered_difference-in-differences}
    \quad log (Y_{ct})=\beta_{0}+\beta_{1} log (S_{ct})+ \lambda_t+\psi_c + \epsilon_{ct}
\end{equation}

Where $log(Y_{ct})$ is a general outcome variable that is indexed by county $c$ and year $t$. $log (S_{ct})$ represents charging the log transformed amount of station subsidies indexed by county, $c$, and year, $t$. $\beta_{1}$ indicates the effect of $log (S_{ct})$ on $log(Y_{ct})$.\footnotemark

\footnotetext{A
positive and significant $\beta_{1}$ suggests that increasing the amount of charging station subsidies exerts a positive effect on the charging station supply, while a negative and significant $\beta_{1}$ indicates that increasing the amount of charging stations decreases charging station supply.} 

Given counties adopted charging station subsidies at different times, I allow for year fixed effects, $\lambda_t$, and county fixed effects $\psi_c$. I include $\lambda_t$ to non parametrically control for national trends in $log(Y_{ct})$, and $\psi_c$ control for any permanent, unobserved differences across counties. These fixed effects incorporate the conditional independence assumption (i.e parallel trends) which states that in the absence of treatment, unobserved shocks affect all units in the same way.\footnotemark 

\footnotetext{ \ It is further assumed that $E[\epsilon_{ct} = 0]$}

For the $\beta_{1}$ estimates to identify the causal effects of $log (S_{ct})$ on $log(Y_{ct})$, it is necessary that
the variables excluded from Equation 1 trend similarly between treatment and control counties (”Parallel Trends Assumption”).\footnotemark
\footnotetext{ \ This assumption would be violated, for example, if public attitudes about EVs and charging station shift differentially in treated counties relative to control counties.} In this log-log model, otherwise known as a constant elasticity model, $\beta_{1}$ reveals that a 1\% change in $log (S_{ct})$ causes a $\beta_{1}$\% change in $log (Y_{ct})$, independent of the initial subsidy amount. This is an important consideration because it allows generalization of $\beta_{1}$ to counties that may have vastly different initial subsidy amounts.

Using Equation (1), I present estimates of treatment subgroups that have high population, high GDP, and are designated as rural or disadvantaged, respectively.\footnote{See notes in Table 4 for further details on these data subsets.} 

In the difference-in-differences estimation technique, I include GDP to control for shocks and trends that affect the economic activity of counties, such as business cycles, changes in regulations and laws, and long-term trends in income distribution. I further include population to control for population  characteristics that may differentially shape EV and charging station attitudes across counties. I allow county level clustering of errors in estimating equation (1), i.e. allowing for correlation in the error terms over time within counties, because subsidy adoption is a county level decision.

While the parallel trends assumption is commonly used by applied economics researchers, it is strong and frequently violated (Abadie, 2005). For example, Ashenfelter (1978) find participants in job training programs have declining earnings prior to the job training treatment (i.e Ashenfelter's dip), thereby violating the parallel trends assumption. Practically, Ashenfelter's dip means that if a participant is already experiencing a decline in the outcome of interest prior to the treatment, then it is difficult to discern the magnitude of the post treatment period change in the outcome variable attributable to the treatment and the portion of the change that is induced by unobserved extraneous factors. I test if the parallel trends assumption in Equation (1) is valid through a generalized event study and the alternative event study specification proposed by Sun and Abraham (2021).

\subsection{Event Study}

The event study analysis allows me to assess the timing of the effects and to check whether trends are parallel before subsidy adoption. I follow Schmidheiny and
Siegloch (2019) in proposing to bin effect window endpoints. The endpoints $-4+$ and 3+ sum up all
maximum lag and lead events beyond the window. I denote the year of switching from
zero subsidies to positive subsidies (i.e subsidy adoption) by event time $\tau = 0$. The generalized event study estimation equation takes the following specification:

\begin{equation}\label{Event_Study_B}
log(Y_{c t})=\beta_{0} + \sum_{\tau=-2}^{\tau=-4+} \gamma_{\tau} \mathbbm{1}(t = E_c+\tau) +\sum_{\tau=0}^{\tau=3+} \delta_{\tau}\mathbbm{1}(t = E_c+\tau)+\lambda_t+\psi_c + \epsilon_{c t}
\end{equation}

Where $log(Y_{ct})$ is a general outcome variable that is indexed by county $c$ and year $t$. $\lambda_t$ are year fixed effects and $\psi_c$ are county fixed effects. $\epsilon_{c t}$ is idiosyncratic noise. The dummy variable $\mathbbm{1}(t = E_c+\tau)$ indicates whether a county has subsidies in a given year for each event year $\tau$ relative to the year of charging station subsidy adoption, where $\tau$ is normalized
to equal zero in the year that a county adopts subsidies; $\tau$ ranges from -4 through 3,
which covers the full range of $\tau$ values. The estimate at $\tau$ = -4 represents the binned value from $\tau$ = -4 to $\tau$ = -12 and the estimate at $\tau$ = 3 indicates the binned value from $\tau$ = 3 to $\tau$ = 12. The omitted reference year is $\tau = -1$, for which all estimates are set equal to zero. For counties that never adopt charging station subsidies,
all $\mathbbm{1}(t = E_c+\tau)$ are set equal to zero. 

The $\gamma_{\tau}$'s and $\delta_{\tau}$’s are the parameters of interest because they report the annual mean of $log(Y_{ct})$ in event time, after adjusting for year and county fixed effects, and the set of controls. An appealing feature of this event study design is that because counties adopted charging station subsidies in different calendar years, it is possible to separately identify the $\gamma_{\tau}$'s, $\delta_{\tau}$’s and the year fixed effects $\lambda_t$.


Although it is common practice to report event study coefficients for the full range of $\tau$ values à la Equation (2), it is possible that non null coefficients in pre-treatment years are not indicative of anticipatory treatment effects, but rather contamination from significant post treatment period effects; recent literature demonstrates that when the effect found from Equation (2) is heterogeneous, it does not represent the weighted average of cohort specific effects and is
biased if these effects vary over time (e.g., Borusyak and Jaravel, 2017; Abraham and Sun, 2021; Athey and Imbens, 2018; de Chaisemartin and D’Haultfœuille, 2018; Goodman-Bacon, 2021; Brzezinski et al., 2020; Chan and Kwok, 2022; Greenstone and Nath, 2020; Strezhnev, 2018; Imai and Kim, 2020). To address this concern, I conduct two additional robustness checks: (1), I re-estimate the event study restricted to the pre-treatment period, and (2), I reproduce the
event-study using the interaction-weighted estimator recommended by
Abraham and Sun (2021). One important feature of the Sun and Abraham design is that it uses a binary independent variable as opposed to the continuous variable used in the staggered difference-in-differences and event study models.

I conduct the robustness check in (1) through an event study design illustrated by the following equation:

\begin{equation}\label{Event_Study_A}
log(Y_{c t})= \beta_{0} + \sum_{\tau=-3}^{\tau=-4+} \delta_{\tau} \mathbbm{1}(t = E_c+\tau) + \delta_{-1}\mathbbm{1}(t = E_c+\tau) + \lambda_t+\psi_c + \epsilon_{c t}
\end{equation}

Equation (3) is a modified version of Equation (2) that restricts to pre-treatment periods since non null pre-trends may be driven by heterogeneous cohort-specific effects. The only differences between Equation (3) and Equation (2) are that in Equation (3), $\tau$ ranges from -4 to -3, with the value at $\tau$ = -4 representing the binned value from $\tau$ = -4 to $\tau$ = -12, and the omitted reference year is $\tau = -2$, for which all coefficients are set equal to zero. $\delta_{-1}$ represents the event study coefficient for $\tau = -1$.

\subsection{Dynamic Treatment Effects}

Recent literature on using traditional two-way fixed effect models of panel data with staggered treatment timing in difference-in-differences estimations caution against interpreting the results as an average treatment effect (Callaway \& Sant’Anna,  2021;
Goodman-Bacon, 2021; Sun \& Abraham, 2021). Sun and Abraham (2021) propose an alternative interaction weighted event study estimation that addresses heterogeneous cohort specific treatment effects. The reasoning for using the Sun and Abraham correction is twofold. First, the general event study specification in Equation (2) does not weight treatment effects proportional to cohort size. Second, in a typical event study approach, anticipatory effects in a pre-treatment relative year, may not suggest violation of parallel trends but rather reflect the true effect of the policy in the post treatment period. Sun and Abraham (2021), with their three part approach, resolve both of these concerns. Under their approach, I estimate the following:  

\begin{equation}\label{Sun and Abraham}
log(Y_{c t})=\beta_{0} + \sum_{e} \sum_{\tau} \rho_{\tau, e \in E} * \mathbbm{1}\left\{E_{c}=e\right\} * D_{c \tau}+\lambda_t+\psi_c+\epsilon_{c t}
\end{equation}

Where $E_{c}$ denotes the year county $c$ adopts charging station subsidies and $E$ denotes the set of all years in which at least one county adopted charging station subsidies. e represents the year of the event (i.e when a county adopts charging station subsidies). The $E_{c}$ indicator is used in addressing contamination of pre-trends from heterogeneous cohort-specific treatment effects. The variables $D_{c \tau}$ are distinct indicators for each event year $\tau$ relative to charging station subsidy adoption, where $\tau$ is normalized
to equal zero in the year that a county adopts subsidies; $\tau$ ranges from -12 through 12, which spans the full range of $\tau$ values.  $log(Y_{c t})$ is a general outcome variable indexed by county $c$, and year $t$. $\tau$ represents event time. $\psi_c$ indicate county fixed effects, and $\lambda_t$ denote year fixed effects. $\beta_{0}$ is a constant and $\epsilon_{c t}$ represents idiosyncratic noise. 

I take a weighted average across cohort-years to aggregate the $\rho_{\tau, e}$ 's to $\rho_{\tau}$. For example, given $\tau=1$, suppose I have a total of 5 observations in our data set, of which 3 are for states who adopted charging station subsidies in 2012 and 2 are for states who adopted charging station subsidies in 2016. Then $\rho_{\tau=1}=\frac{3}{5} * \rho_{\tau=1, e=2012}+\frac{2}{5} * \rho_{\tau=1, e=2016}$. 

The accuracy of the predicted treatment effect can be increased if the $\rho_{\tau}$ estimate is converted from log points to percentage points. Under the central assumption that the error term, $\epsilon_{c t}$, is not correlated with $\mathbbm{1}\left\{E_{c}=e\right\} * D_{c \tau}$, the exact percentage change in $log(Y_{c t})$ for $\Delta \mathbbm{1}\left\{E_{c}=e\right\} * D_{c \tau}$ is evaluated:

\begin{equation}\label{Log Conversion Equation}
\% log(Y_{c t})=\left(\exp \left(\left(\Delta \mathbbm{1}\left\{E_{c}=e\right\} * D_{c \tau}\right)\beta_1\right)-1\right) \times 100
\end{equation}

Where $\Delta \mathbbm{1}\left\{E_{c}=e\right\} * D_{c \tau}$ is the change in the dummy variable $(=1-0)$ and $\rho_{\tau}$ is the treatment effect expressed in log points. The percentage change in $log(Y_{c t})$ approximated by $\rho_{\tau}$ is always a lower bound of the actual percentage change in $log(Y_{c t})$ The discrepancy between the approximation and exact percentage change in $log(Y_{c t})$ caused by $\Delta \mathbbm{1}\left\{E_{c}=e\right\} * D_{c \tau}$ rises exponentially as the $\rho_{\tau}$ coefficient increases.\footnote{The percentage change in $log(Y_{c t})$ approximated by $\rho_{\tau}$ is always a lower bound of the actual percentage change in $log(Y_{c t})$.}


\section{Results}
To investigate the effect of charging station subsidies on charging station supply, I estimate causal effects from two models: a staggered difference-in-differences equation (Equation 1), and the alternative interaction-weighted estimator proposed by Sun and Abraham (2021) (Equation 4). I further conduct an event study restricted to the pre-treatment period as a robustness check to probe the validity of the parallel trends identifying assumption.

\subsection{Effect of Charging Station Subsidies on Supply}
In this section, I examine the effects of the subsidy on charging station supply, considering various differences in county profile. I use the staggered difference-in-differences model in Equation (1).

Table 3 displays point estimates for the baseline difference-in-differences specification with county and year fixed effects and GDP and purchase price subsidy controls. As seen in Table 3, a 100\% increase in the magnitude of subsidies increases supply by 2.5\% (std. err. 0.008). GDP has a positive effect on all the dependent variables of interest, while purchase price subsidies have a positive effect on DC stations and BEV sales. Purchase price subsidies, which directly reduce the price of EVs, do not have a direct effect on charging stations. Meanwhile, GDP has a positive effect on stations and sales — regions with greater wealth tend to have greater EV adoption and charging station availability. These findings are consistent with Figure 3 which shows that purchase price subsidies and GDP are positively correlated with charging station subsidies, charging station supply, and EV sales. The negative coefficient of purchase price subsidies on PHEV sales could be due to the possibility that purchase price subsidies target BEVs more than PHEVs.  

Table 4 reports difference-in-differences estimation results for six different specifications from the staggered difference-in-differences design where the independent variable is DC station subsidies. Given the log-log specification, the reported coefficients can be interpreted as elasticities, and standard errors are robust against heteroskedasticity and autocorrelation due to clustering at the county level. 

Row 1 is the baseline estimation with county and year-fixed effects. Row 2 adds two additional variables — GDP and purchase price subsidies — to control for observed differences in economic activity and alternative EV incentives, respectively, that may affect charging station supply and sales outcomes. 

As shown in the baseline specification in Column 1, a 100\% increase in the magnitude of DC subsidies increases the supply of DC stations by 2.6\% (std. err. 0.009). In the baseline + controls specification, the effect is slightly smaller: a 100\% increase in the magnitude of subsidies increases supply by 2.5\% (std. err. 0.008). These positive coefficients are consistent with Figure 2 which documents a strong correlation between subsidies and station supply and between subsidies and sales. The similar coefficients between the baseline and control specification suggests that substantial omitted variable bias is unlikely. This central finding demonstrates that charging station subsidy supply is elastic in respect to charging station subsidies. This finding affirms the theoretical expectation that charging station subsidies are successful in increasing supply, and suggests that policymakers in other states should consider implementing charging station subsidies if they aim to stimulate their charging station market.\footnote{Although I demonstrate in Section 5.3 that the relationship between EV sales and charging station subsidies lacks causal interpretation, it is possible that such subsidies indirectly increase EV sales through expanding charging access — a topic that future research should investigate.} 

In the baseline + controls specification, I find the elasticity of PHEV sales (BEV sales) in respect to subsidies to be 0.021 (0.016), suggesting that a 100\% increase in subsidies is associated with a 2.1\% (1.6\%) increase in PHEV sales (BEV Sales). Given that these elasticities are overestimates, this finding reinforces the conclusion of Luo (2021): a 100\% increase in the Clean Vehicle Credit (i.e \$7,500 federal EV tax credit) is associated with a 0.432\% increase in EV sales, which is 79.4\% less than than the estimate on BEV sales (2.1\%) and 73\% less than the estimate on PHEV sales (1.6\%).\footnote{The elasticity in Luo (2021) is calculated by dividing the subsidy’s effect on EV sales, 43.2\%, by 100.} Even though the elasticities are overestimates, they are still substantively larger than that found in Luo (2021) to suggest charging station subsidies may be more effective than purchase price subsidies at increasing EV sales, which is consistent with findings from Li et al. (2017) and Cole et al. (2021).

\subsection{Heterogeneity}

I implement the specification in Row 2 which includes the control variables on various subsets of the data to determine if subsidy effects vary based on county characteristics. In Table 4, Rows 3-6 include the same controls and county and year fixed effects in Row 2 but differ in the following ways: Row 3 subsets to counties whose population is greater than the median California county population; Row 4 only considers counties classified as disadvantaged in 2020; Row 5 subsets to counties classified as rural in 2020 in accordance with the guidelines from the California State Association of Counties; and Row 6 subsets to counties with GDP higher than the median GDP across all California counties in 2020.\footnote{See data dictionary for more detailed descriptions on selection criteria.}

In Rows 3-6, I find that the coefficients on DC stations — 0.021 (std. err. 0.009), 0.019 (std. err. 0.010), 0.009 (std. err. 0.014), 0.021 (std. err. 0.008) — respectively, are smaller than the coefficient of 0.026 in the baseline + controls specification; in a pattern common to all the Column 1 Row 3-6 coefficients, the value of $\beta_{1}$ is either smaller in magnitude or less significant than the control specification.\footnote{In a pattern common to all specifications in Tables 3 and A1, I find null coefficients on Tesla Destination (Level 2) and Supercharger (DC) stations. This finding is consistent with my expectation that Tesla is less responsive to subsidies due to the different incentives it faces compared to conventional charging station suppliers. Tesla could free ride in that they may claim government subsidies even when the subsidies did not induce them to build new stations. To address this shortcoming, policymakers should consider excluding Tesla stations from subsidy programs to increase cost efficiency.} Although I expect that the $\beta_{1}$ coefficients on the subsetted groups are greater than that on the controls or baseline specification, my results show the opposite. One possible explanation is that consumers in areas with high population density may express a more unfavorable attitude towards electric vehicles generally, making constructing charging stations particularly unprofitable. Another possibility is that consumers in these regions may be so impoverished that consumers may not be able to purchase EVs whether they have access to charging or not. However one caution is that the results on the heterogeneous subgroups are subject to statistical uncertainty due to larger standard errors. All of the subgroups by definition rely on a smaller sample size. Accordingly, the specifications in Rows 3-6 have a smaller degree of statistical significance compared to the baseline + controls specification. Indeed, the Column 1 coefficient among Rows 3-6 with the least statistical significance — the urban specification — also has the least number of observations (111). Given this consideration, I rely more on the control specification for a generalized elasticity result than the subsetted groups due to its relative statistical strength. 

As seen in the baseline + controls specifications, I document positive effects of DC subsidies on PHEV and BEV sales. Among the specifications in Rows 3-6, however, the only significant coefficient is on PHEV sales in the high population specification. 

\subsection{Robustness Checks}
\subsubsection{Event Study} 
Table 5 reports coefficients on the interaction term between treatment status and relative years in the pre-treatment period and the relative post-treatment year $\tau = -1$. A county is defined to be treated if it adopts DC subsidies in the period 2010 - 2022 and untreated if it does not. The omitted reference year is $\tau = -2$. As shown in Columns 1 and 2, I document null pre-trends in DC stations and Tesla Supercharger stations. This observed lack of anticipatory effects supports the parallel trends assumption in the staggered difference-in-differences specification (see Equation 1).\footnote{I drop the elasticity interpretation in the event study and follow up Sun and Abraham correction since these models define a binary rather than continuous treatment variable.}

Perhaps more interestingly, however, I document anticipatory effects of Log(Sales), Log(PHEV Sales), and Log(BEV Sales) for the binned value $\tau = -4$ to $\tau = -12$ and the relative event year $\tau = -3$. The pre-treatment period coefficients in Columns 3-5 are all significant and positively trending: in Log(Sales) the binned value coefficient of -0.23 (std. err. 0.10) increases by 0.10 to -0.13 (std. err. 0.04); in Log(PHEV Sales) the binned value coefficient of -0.14 (std. err. 0.08) increases by 0.03 to -0.11 (std. err. 0.05); and in Log(BEV Sales) the binned value coefficient of -0.43 (std. err. 0.18) increases by 0.26 to -0.17 (std. err. 0.010).\footnote{For example, the coefficient of -0.23 (std. err. 0.10) on the binned value in Log(Sales) means that EV sales decreased by 23\% in the period 4 to 12 years before the enactment of the subsidy.}  The existence of significant pre-trends demonstrates that failure to properly model EV sales dynamics, or the propensity to adopt charging station subsidies based on past EV adoption trends, will lead to biased OLS estimates of charging station subsidies on EV sales. In particular, the positive trend in sales measures reveals that counties that adopt charging subsidies tend to have growing EV sales even prior to subsidy adoption, and further suggests that sales measures may be correlated with unobserved covariates. Thus, the regression results in Table 5 are consistent with Table 2, which reveals that treated counties on average have significantly more EV sales than untreated counties at time of subsidy adoption. This violation in pre-trends is analogous to the one popularized in “Ashenfelter’s dip” (Ashenfelter, 1978). 

Future research that examines the effect of charging station subsidies on EV sales should account for endogeneity in EV sales, for example through an instrumental variable approach. If policymakers act rationally in incorporating knowledge of under investment in charging station infrastructure in their policy decisions, one would expect counties to implement subsidies when EV sales are low or negatively trending so as to rectify the market failure. However, I observe counties that adopt charging subsidies already experience rising EV sales trends even before subsidy adoption.

\subsubsection{Sun and Abraham Correction for Staggered Treatment Timing}
Although the general event study specification in Equation (3) restricts to the pre-treatment period, significant observed pre-trends could actually reflect the true post-treatment effect of the subsidy. In Table 6, I reproduce the event-study using the interaction-weighted estimator recommended by Abraham and Sun (2021) to address concerns of heterogeneous cohort specific effects. Consistent with Table 5, I document null pre trends in DC stations and Tesla superchargers, further supporting the parallel trends identifying assumption in the staggered difference-in-differences model. Interestingly, the effect of DC subsidies on DC stations is positive and statistically significant after $\tau = 1$ and increases by 0.14 from 0.36 (std. err. 0.15) in $\tau = 2$ to 0.50 (std. err. 0.17) in the binned value $\tau = 3$ to $\tau = 12$; using the conversion from log points to percentage points in Equation (5), I find the enactment of DC subsidies increased DC station supply by 36\% two years after enactment and by 50\% between 3 and 12 years after enactment. One reason for this lagged effect is that public charging stations take a considerable amount of time to plan, construct, and become fully operational: theoretically, a charging station whose construction process begins at $\tau = 0$ may not open to the public until $\tau = 2$, reflecting the lagged effect. Another reason could be that investors do not immediately respond to the charging station subsidies. Consistent with theoretical expectation, there are null coefficients throughout the full range of $\tau$ values for Tesla Supercharger stations. 

Similar to Table 5, I find significant positive pre-trends for the binned values from $\tau = -3$ to $\tau = -12$ in aggregate sales, PHEV sales, and BEV sales — -0.21 (std. err. 0.06), -0.20 (std. err. 0.07), and -0.37 (std. err. 0.09) — respectively. I do not find significant post-treatment effects on sales measures. Moreover, the positive coefficient in the post-treatment period on Log(PHEV Sales) suggests that growing sales is a continuation of faster sales growth in treated counties. Both the event study and Sun and Abraham estimation exhibit pre-trends in EV sales (see Tables 4 and 5, respectively), suggesting that any significant documented effects of charging station subsidies on EV sales are likely driven by OVB.

Figure 4 provides a graphical depiction of how the point estimates on the event study model restricted to the pre-treatment period differ from those on the Sun and Abraham correction. In Panel A (Panel B), I observe a significant negative (null) pattern in pre-trends in both the generalized event study and the Sun and Abraham correction. However, the significant negative pre-trend in Panel A does not necessarily invalidate causal interpretation because the negative trend in reverses immediately to a positive trend following subsidy adoption, which seems to suggest a meaningful causal effect of the subsidy on station supply. Consistent with my prior findings that treated counties have a greater EV presence prior to subsidy adoption, the observed negative pre-trends may still suggest rising nominal charging station counts. Since trends in treated counties are relative to control counties, it is possible that negative pre-trends indicate charging station supply grows at a slower rate in counties that adopt subsidies than in counties that do not. 

In Panel A, although the event study does not yield a significant post-treatment effect, the Sun and Abraham correction does reveal a significant effect in $\tau = 2$ — the subsidy increases Level 2 station supply by 25\% two years following subsidy adoption. In contrast to Panel A, Panel B displays null pre-trends for both the event study and Sun and Abraham specifications and demonstrates that they move more in tandem throughout the full range of $\tau$ values. This finding validates the parallel trends condition, which allows me to assign elasticity interpretations to the difference-in-differences coefficients in the post-treatment period. Moreover, the post-treatment period coefficients are significant and positive beginning at $\tau = 2$ in the event study and Sun and Abraham correction.

\section{Conclusion}
Although the central role of charging stations in EV adoption has received growing policy attention in recent years, it is unclear to what extent charging station subsidies stimulate charging station supply. The study provides to my knowledge the first empirical analysis of the effect of charging station subsidies on charging station supply using the approach recommended by Sun and Abraham (2021). Using this novel methodological approach, I estimate the effect of DC charging station subsidy adoption on DC station supply to be 0.36 (std. err. 0.15) and the effect of Level 2 charging station subsidy adoption on Level 2 station supply to be 0.22 (std. err. 0.11), two years following subsidy adoption. Thus, adoption of DC (Level 2) charging subsidies increases DC (Level 2) increases charging station supply by 36\% (24.6\%) two years after subsidy adoption. This finding suggests that charging station subsidies, at least on the county level in California, are highly effective at increasing charging station supply. If Level 1 stations become more commonplace in the future, researchers should likewise examine the effect of Level 1 subsidies on Level 1 stations. In light of recent legislation, studies that examine the impact of federal charging station grants on charging station supply could provide insight into the effectiveness of alternative methods of government intervention in the charging station market.\footnote{See, for example, the 2021 Infrastructure Investment and Jobs Act, which “invest[s] \$7.5 billion to build out a national network of EV chargers in the United States” (The White House, 2021).} The effects of the subsidy on charging station supply are not substantially different across various subgroups. In particular, the effects of the subsidy in denser regions do not appear stronger than those in less dense regions. I also do not find that disadvantaged counties benefit more from these subsidies than non-disadvantaged counties.\footnote{While these results seem to reject economic theory, due not because of their larger standard errors; by definition, these subgroups rely on a smaller sample size. One method to increase the sample size in the future is to draw from a longer observation period. Another method is to use county level data from states beyond California once they establish a greater charging station presence.}

With respect to the event study and Sun and Abraham models, one limitation is that most treated counties adopted subsidies late in the observation period. For these ‘late adopters,’ the availability of post treatment effects may be limited.\footnote{For instance, a county that adopted subsidies in 2021 observes only the one year relative effect (calendar year 2022).} As the observation period naturally increases over time, follow up studies should use data from more relative years. Such studies should also seek to determine if, consistent with my findings, charging station subsidies generally have greater effects around the first or second relative year. On the same token, it may be insightful to examine if the effect becomes insignificant some years after the introduction of the subsidy.

Beyond the causal effect of charging subsidies on charging station supply, my results offer some insights for future research and policy design. Most significantly, the high elasticity of charging stations in respect to charging station subsidies suggests policymakers can consider increasing charging station subsidies to promote charging station supply. In particular, they may enact more subsidies if they wish to address under-investment in charging station infrastructure. 

Considering significant pre-trends in sales, the question arises of what portion of the observed effect is due to the subsidy and what portion is due to existing investment being directed into counties with the subsidy. Theoretically, investors could have aimed to invest a given amount in charging stations and choose to invest in counties where they can gain additional profits from the subsidy — in this situation, the subsidy does not induce these agents to invest. Although this study addresses this issue through the use of year-fixed effects, future researchers should run additional tests for private charging stations owned by businesses without geographical flexibility to investigate further. The finding that subsidy adoption is responsive to EV sales suggests future research investigating the direct effect of charging station subsidies on EV sales can consider using an instrumental variable approach. More specifically, my discovery of rising EV sales prior to subsidy adoption motivates several fruitful paths for further investigation. First, researchers should examine if rising EV sales pre-trends among treated counties is permanent or transitory. Second, given my findings are limited to the scope of California counties, future studies should investigate if states outside California or other countries similarly experience positive and growing EV sales pre-trends. The results from such an investigation could present significance for evaluating the effectiveness of national level charging subsidy policy. Third, I found that policymakers implement charging station subsidies when EV sales are growing. However, it is unclear if this is also the case for purchase price subsidies; future research should build on my findings by examining if there are significant pre-trends in the relationships between purchase price subsidies and EV sales.

While it is expected that government entities will intervene when the market fails to produce charging stations at the socially optimal quantity, I find that counties that adopt charging subsidies tend to have rising EV sales trends prior to subsidy adoption compared to never adopters. This implies that the government in counties with stable or declining EV sales trends — those most affected by the under-investment in charging station infrastructure — don’t adopt the subsidies they need to stimulate charging station supply and EV adoption to their socially optimal quantity. In other words, these governments tend to invest in regions that already have a developed EV presence as opposed to those without. Together with the strong effect of charging station subsidies on station supply, the responsiveness of subsidy adoption to prior sales trends argues for alternative subsidy designs to ensure an equitable distribution of charging station subsidies.


\clearpage
\nocite{*}

\bibliographystyle{aea}
\bibliography{references.bib}

\clearpage

\begin{figure}[hp]\caption[]{}
\centering
\begin{minipage}{.7\linewidth}
\includegraphics[width=\linewidth]{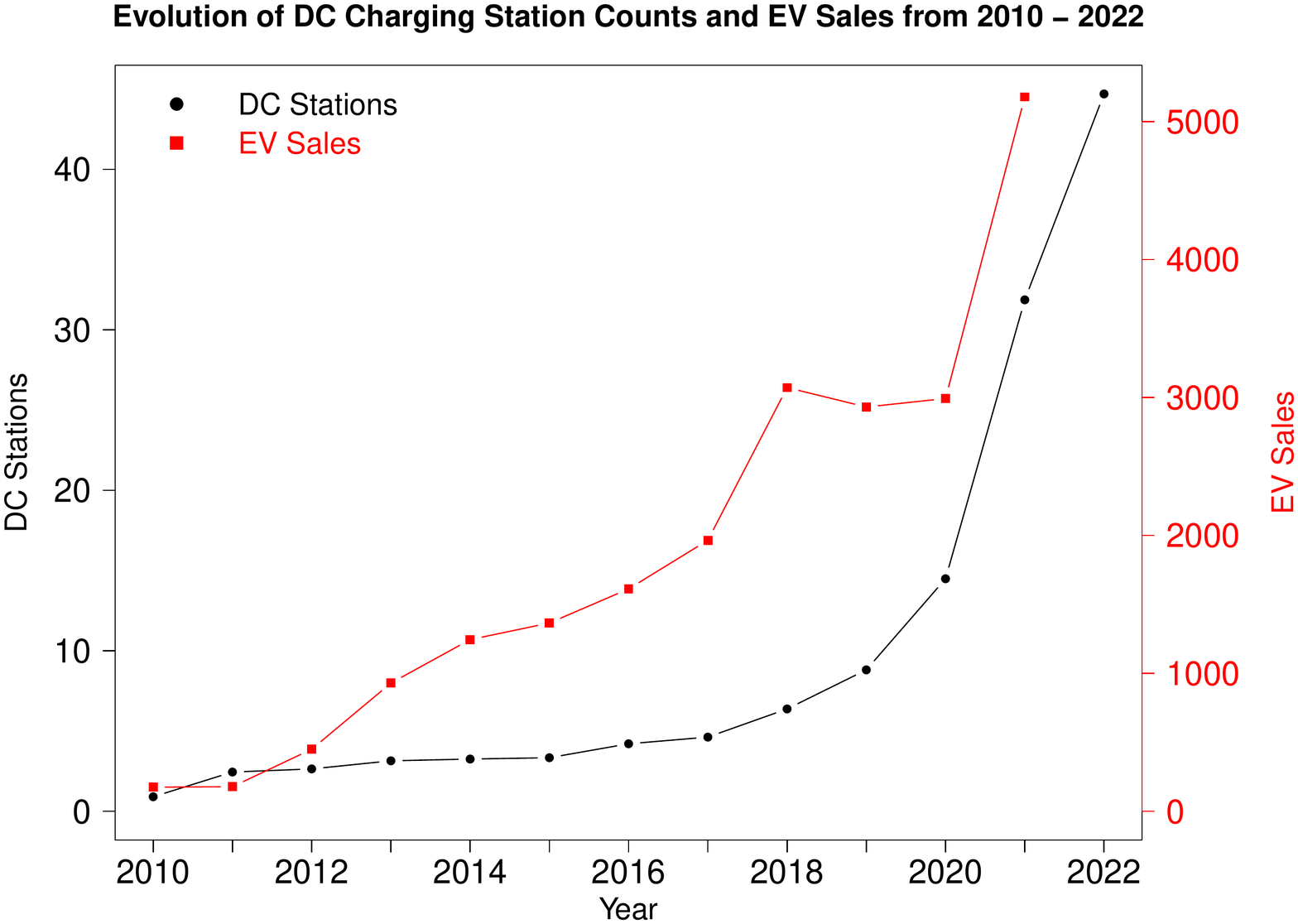}
\small
\emph{Notes:} The sample consists of time series data of DC charging stations from 2010 to 2022. Individual data points represent average DC charging stations across all observed California counties. DC station counts
are retrieved from the National Renewable Energy Laboratory.  \par
\end{minipage}
\end{figure}






\clearpage
\begin{figure}[htp]
\centering
 \caption{}
 \centering
   \includegraphics[width=\linewidth]{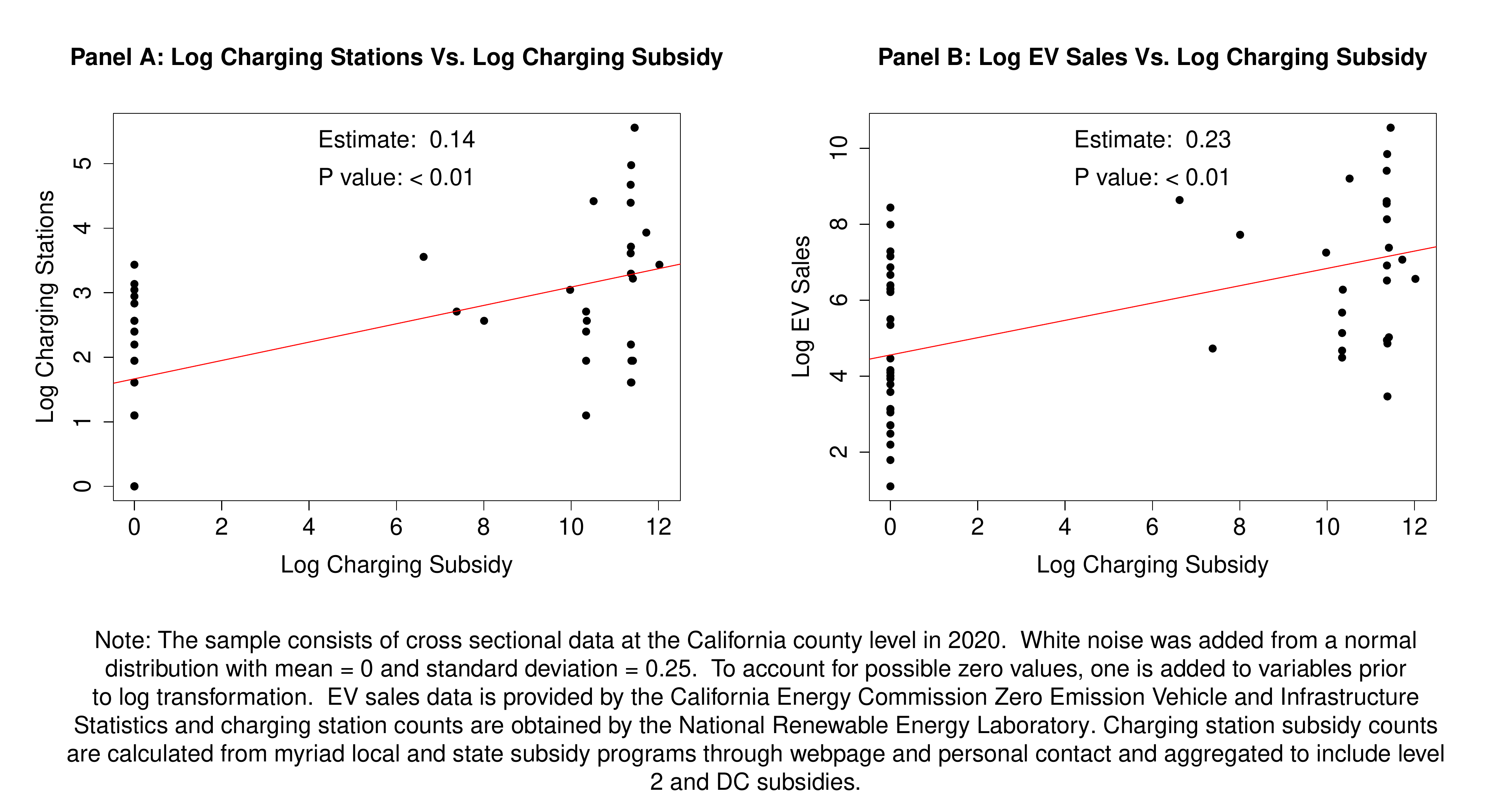}
\end{figure}

\begin{landscape}

\clearpage
\begin{figure}[htp]
 \centering
   \centering
    \caption{}
   \includegraphics[scale = 0.55]{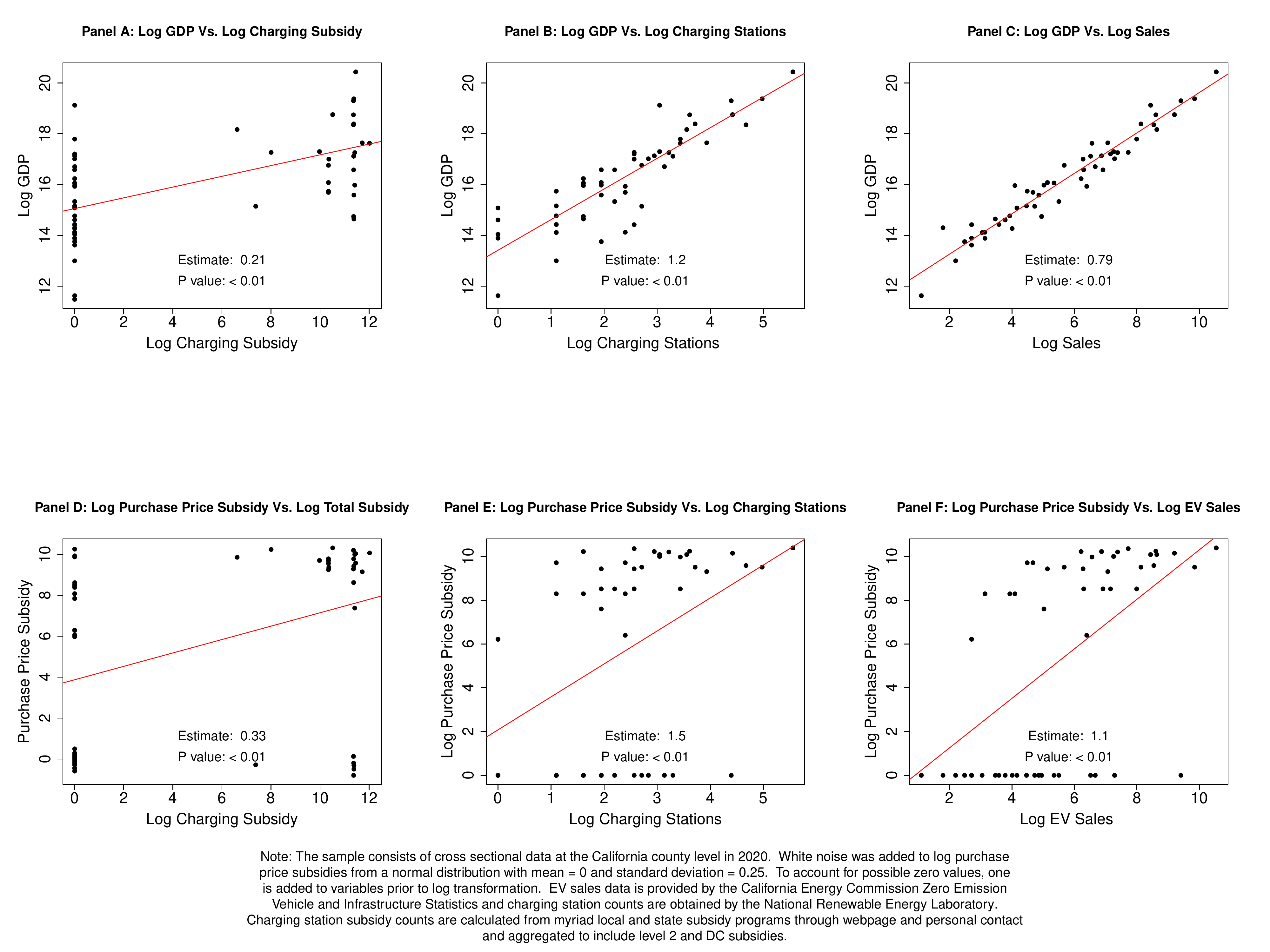}
\end{figure}
\end{landscape}

\begin{figure}[hp]\caption[]{}
\centering
\begin{minipage}{.7\linewidth}
\includegraphics[width=\linewidth]{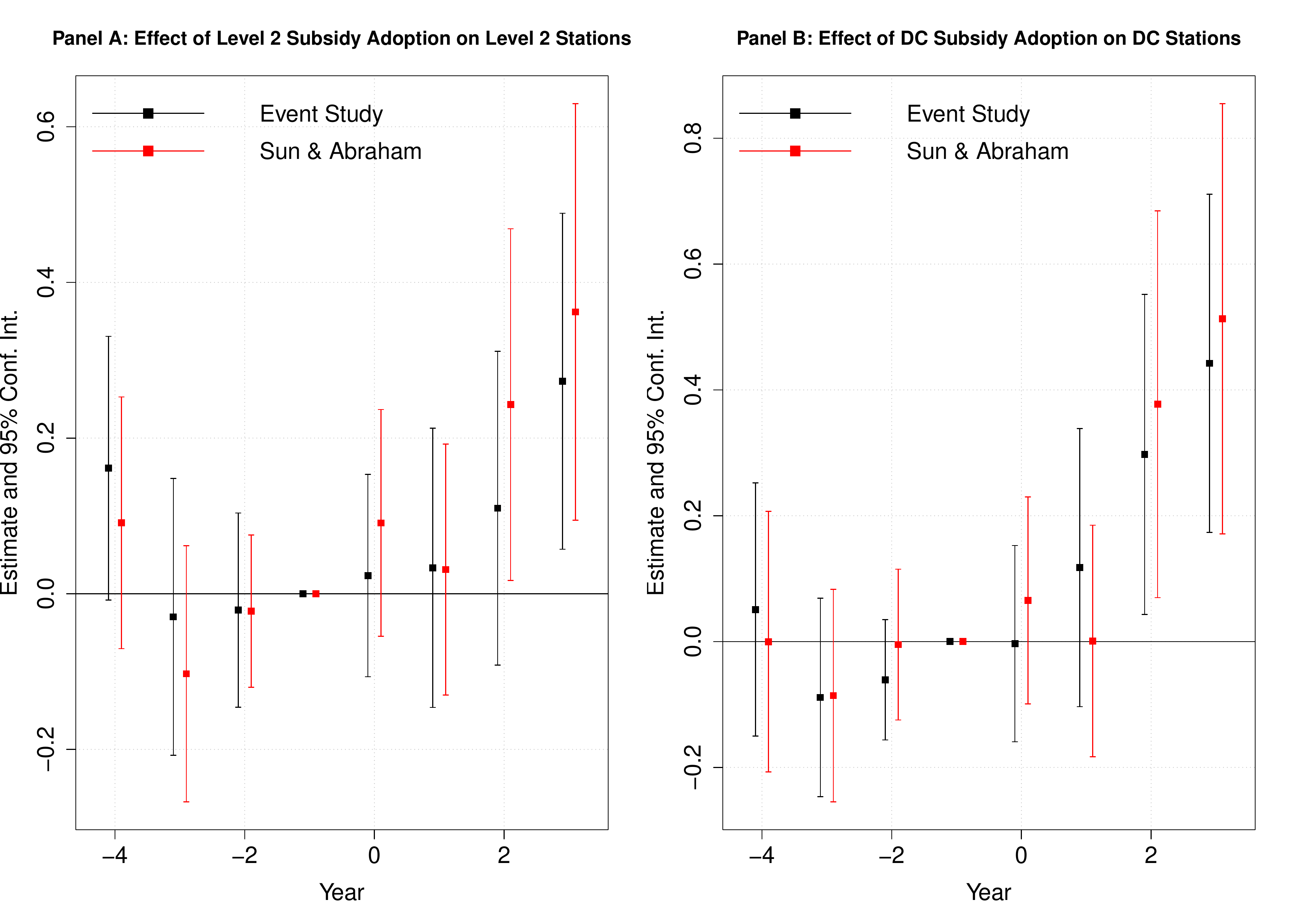}
\small
\emph{Notes:} The red lines display difference-in-differences coefficients and corresponding 95\% confidence intervals for the alternative specification with staggered treatment timing using an “interaction-weighted” estimator proposed by Abraham and Sun (2021), while the black line displays coefficients and their corresponding 95\% confidence intervals from the event study specification in Equation (2). The red lines corresponds to 95\% confidence intervals of $\rho_\tau$ for $\tau$ = -4 to $\tau$ = 3. For each treatment cohort (defined by whether the county adopts charging subsidies) in the event study (Sun and Abraham) specification I estimate Equation (4) against the control group of never treated counties. The value at $\tau$ = -4 represents the binned value from $\tau$ = -4 to $\tau$ = -12 and the value at $\tau$ = 3 indicates the binned value from $\tau$ = 3 to $\tau$ = 12 for the event study specification in Equation (4) that allows for cohort-year interactions with the $\rho_\tau$’s. Standard errors are clustered at the county level.\par
\end{minipage}
\end{figure}

\newpage
\clearpage
\begin{table}
\small
\caption{\label{tab:year}Number of Counties Adopting Subsidies by Year}
\centering
\begin{threeparttable}
\begin{tabular}[t]{lrrrrrrrrrrrrr}
\toprule
  & 2010 & 2011 & 2012 & 2013 & 2014 & 2015 & 2016 & 2017 & 2018 & 2019 & 2020 & 2021 & 2022\\
\midrule
DC Adopters & 0 & 0 & 0 & 0 & 0 & 0 & 1 & 2 & 11 & 6 & 2 & 13 & 0\\
Level 2 Adopters & 0 & 1 & 0 & 1 & 1 & 1 & 1 & 3 & 9 & 6 & 3 & 12 & 0\\
\bottomrule
\end{tabular}
\begin{tablenotes}
\item \textit{Note: } 
\item A county X is defined to adopt in year Y if it has a positive amount of charging subsidies in year Y and zero subsidies in year Y - 1. This treatment definition avoids duplicate observations since treated counties may increase their subsidy amount multiple times throughout the observation period. There are no recorded counties that adopted subsidies in 2022 because complete data was unavailable at the time of retrieval. Data on cumulative county-year subsidy counts by station type is sourced from manual compilation of subsidy amounts from various county and state web pages. See legal appendix for original sources.
\end{tablenotes}
\end{threeparttable}
\end{table}

\begin{table}
\caption{\label{tab:CountyChar}Charging Stations, Sales, and County 
Characteristics in 2020: Descriptive Statistics}
\centering
\begin{threeparttable}
\begin{tabular}[t]{lcccccc}
\toprule
  & Untreated & Treated & Difference & P Value & \makecell[c]{N: \\ Treated} & \makecell[c]{N: \\ Untreated}\\  
                   & \makecell[c]{(1)}          & \makecell[c]{(2)}                             & \makecell[c]{(3)}          & \makecell[c]{(4)} & \makecell[c]{(5)} & \makecell[c]{(6)}\\       
\midrule
\addlinespace[0.3em]
\multicolumn{7}{l}{\textbf{Charging Station Counts}}\\
\hspace{1em}Log Level 2 & 2.58 & 4.25 & -1.68 & 0.00 & 26 & 25\\
\hspace{1em}Log DC & 1.23 & 2.56 & -1.33 & 0.00 & 26 & 25\\
\hspace{1em}Log Tesla Super Charger (DC) & 2.16 & 3.76 & -1.60 & 0.00 & 26 & 25\\
\hspace{1em}Log Tesla Destination (Level 2) & 2.25 & 3.32 & -1.07 & 0.03 & 26 & 25\\
\hspace{1em}Log Total & 2.80 & 4.42 & -1.62 & 0.00 & 26 & 28\\
\addlinespace[0.3em]
\multicolumn{7}{l}{\textbf{Subsidy Amounts}}\\
\hspace{1em}Log Level 2 & 0.00 & 8.87 & -8.87 & 0.00 & 26 & 30\\
\hspace{1em}Log DC & 0.00 & 9.28 & -9.28 & 0.00 & 26 & 28\\
\hspace{1em}Log Total Charging Station & 0.00 & 10.67 & -10.67 & 0.00 & 26 & 30\\
\hspace{1em}Log Purchase Price & 3.81 & 7.45 & -3.64 & 0.00 & 26 & 30\\
\addlinespace[0.3em]
\multicolumn{7}{l}{\textbf{Sales Measures}}\\
\hspace{1em}Log BEV & 4.02 & 6.67 & -2.65 & 0.00 & 26 & 30\\
\hspace{1em}Log PHEV & 3.32 & 5.75 & -2.43 & 0.00 & 26 & 30\\
\hspace{1em}Log Total & 4.54 & 7.01 & -2.46 & 0.00 & 26 & 30\\
\addlinespace[0.3em]
\multicolumn{7}{l}{\textbf{Market Characteristics}}\\
\hspace{1em}EV Share & 0.03 & 0.05 & -0.01 & 0.03 & 26 & 30\\
\hspace{1em}ICE Share & 0.97 & 0.95 & 0.01 & 0.03 & 26 & 30\\
\addlinespace[0.3em]
\multicolumn{7}{l}{\textbf{Demographic Characteristics}}\\
\hspace{1em}Log Population & 11.09 & 13.29 & -2.20 & 0.00 & 26 & 30\\
\hspace{1em}Log GDP & 15.06 & 17.31 & -2.26 & 0.00 & 26 & 25\\
\bottomrule
\end{tabular}
\begin{tablenotes}
\item \textit{Note: } 
\item The sample restricts to county-year observations in 2020 for all observed counties in California. Sample sizes differ across rows due to missing values. Raw and log transformed means and p-values from a two sided t-test reported. Counties are denoted as treated (untreated) if they do (do not) receive a positive amount of charging station subsidies in the period 2010 - 2022. Negative differences indicate the estimate is higher for treated counties than untreated ones, whereas positive differences indicate that the estimate is higher for untreated counties than treated counties. Tesla destination stations are classified as level 2 stations and are a part of Tesla's proprietary charging network. Tesla super chargers are classified as DC stations and are a part of Tesla's proprietary charging network. To account for possible zero values, one is added to variables prior to log transformation. Values in columns (1) and (2) represent mean values in 2020 and nominal amounts are log transformed. Values in column (3) are calculated by subtracting the mean value for untreated counties from that of treated counties. Column (4) reports p-values for the values in column (3).
\end{tablenotes}
\end{threeparttable}
\end{table}

\begin{table}[htbp]
   \caption{\label{tab:main}Effect of Charging Station Subsidies on Charging Stations}
   \centering
   \begin{threeparttable}
   \begin{tabular}{lcccc}
      \tabularnewline \midrule \midrule
      Dependent Variables:    & \makecell[c]{Log\\(DC)}       & \makecell[c]{Log(Tesla Super \\ Charger Stations)} & \makecell[c]{Log\\(PHEV Sales)} & \makecell[c]{Log\\(BEV Sales)}\\  
      Model:                  & (1)           & (2)                               & (3)             & (4)\\  
      \midrule
      \emph{Variables}\\
      Log(DCS)                & 0.025$^{***}$ & 0.031                             & 0.021$^{***}$   & 0.017$^{**}$\\   
                              & (0.008)       & (0.019)                           & (0.007)         & (0.008)\\   
      Log(GDP)                & 0.432         & 0.123                             & 1.23$^{***}$    & 0.732\\   
                              & (0.551)       & (0.541)                           & (0.455)         & (0.483)\\   
      Log(Purchase Price Sub) & 0.003         & -0.007                            & -0.007          & 0.034$^{***}$\\   
                              & (0.009)       & (0.015)                           & (0.008)         & (0.009)\\   
      Number of Counties      & 56            & 56                                & 56              & 55\\  
      \midrule
      \emph{Fixed-effects}\\
      County                  & Yes           & Yes                               & Yes             & Yes\\  
      Year                    & Yes           & Yes                               & Yes             & Yes\\  
      \midrule
      \emph{Fit statistics}\\
      Observations            & 410           & 410                               & 549             & 554\\  
      Dependent variable mean & 1.3           & 0.60                              & 4.3             & 4.4\\  
      R$^2$                   & 0.95          & 0.77                              & 0.98            & 0.98\\  
      \midrule \midrule
      \multicolumn{5}{l}{\emph{Clustered (County) standard-errors in parentheses}}\\
      \multicolumn{5}{l}{\emph{Signif. Codes: ***: 0.01, **: 0.05, *: 0.1}}\\
   \end{tabular}
   \begin{tablenotes}
   \item The sample consists of data at the county-year level from 2010 - 2022 for all observed counties within California. This model is the same as the baseline + controls specification in Table 4. Standard errors are clustered at the county level. The regression incorporates year and county fixed effects. GDP and population data are retrieved from the California Department of Finance. All charging station counts are derived from the National Renewable Energy Laboratory and charging station subsidy counts and purchase price subsidy counts are retrieved from myriad local and state subsidy programs through webpage and personal contact. GDP and purchase price subsidy counts are control variables.
   \end{tablenotes}
   \end{threeparttable}
\end{table}

\clearpage

\begin{table}
\caption{\label{tab:did1}Effects of Subsidies on Stations and Sales Measures: Staggered Difference-in-Differences}
\centering
\begin{threeparttable}
\begin{tabular}[t]{lllll}
\toprule
  & Log(DC) & \makecell[c]{Log(Tesla Super\\Charger Stations)} & Log(PHEV Sales) & Log(BEV Sales)\\
\midrule
Baseline & 0.026 (0.009) *** & 0.025 (0.015) & 0.018 (0.006) *** & 0.019 (0.007) **\\
Baseline + Controls & 0.025 (0.008) *** & 0.031 (0.019) & 0.021 (0.007) *** & 0.016 (0.008) **\\
High Pop & 0.021 (0.009) ** & 0.027 (0.021) & 0.011 (0.006) ** & 0.002 (0.006)\\
Disadvantaged & 0.019 (0.010) * & 0.018 (0.019) & 0.007 (0.006) & -0.011 (0.007)\\
Urban & 0.009 (0.014) & 0.005 (0.012) & 0.031 (0.019) & 0.019 (0.018)\\
\addlinespace
High GDP & 0.021 (0.008) ** & 0.025 (0.021) & 0.010 (0.006) & 0.003 (0.006)\\
\bottomrule
\end{tabular}
\begin{tablenotes}
\item \textit{Note: } 
\item The sample restricts to an unbalanced panel of county-year observations for all observed counties in California. The time period covered is 2010 - 2022. The table presents the coefficients and standard errors on the treated x post treatment period interaction term. The treatment is defined as a county’s adoption of charging station subsidies. Treated counties adopt charging station subsidies within the time period 2010-2022 and the post treatment period is the period of time after the adoption of charging stations subsidies to 2022. Standard errors are clustered at the county level. Charging station counts and sales measures are log transformed. To account for possible zero values, one is added to variables prior to log transformation. Row 2 adds controls for GDP and purchase price subsidies. Row 3 subsets to counties whose population is greater than the median population among all California counties. Row 4 only considers counties classified as disadvantaged in 2020 in accordance with SB 535. Row 5 subsets to counties classified as rural in 2020 in accordance with the guidelines from the California State Association of Counties. Row 6 subsets to counties with GDP higher than the median GDP across all California counties in 2020. Tesla destination stations are classified as level 2 stations and are a part of Tesla's proprietary charging network. Tesla superchargers are classified as DC stations and are a part of Tesla's proprietary charging network.
\end{tablenotes}
\end{threeparttable}
\end{table}


\begin{table}[htbp]
   \caption{\label{tab:ols1} Effects of Charging Stations on Stations and \\ Sales Measures: OLS Results}
   \centering
   \begin{threeparttable}
   \begin{tabular}{lccccc}
      \tabularnewline \midrule \midrule
      Dependent Variables:                      & \makecell[c]{Log\\(DC)} & \makecell[c]{Log(Tesla Super \\ Charger Stations)} & \makecell[c]{Log\\(Sales)}    & \makecell[c]{Log\\(PHEV Sales)} & \makecell[c]{Log\\(BEV Sales)}\\  
      Model:                                    & (1)     & (2)                               & (3)           & (4)             & (5)\\  
      \midrule
      \emph{Variables}\\
      DC Treated $\times$ Relative Year $=$ -4 to -12  & -0.09   & -0.05                             & -0.23$^{**}$  & -0.14$^{*}$     & -0.43$^{**}$\\   
                                                & (0.11)  & (0.16)                            & (0.10)        & (0.08)          & (0.18)\\   
      DC Treated $\times$ Relative Year $=$ -3  & -0.05   & -0.02                             & -0.13$^{***}$ & -0.11$^{**}$    & -0.17$^{*}$\\   
                                                & (0.09)  & (0.10)                            & (0.04)        & (0.05)          & (0.10)\\   
      DC Treated $\times$ Relative Year $=$ -1  & 0.005   & 0.04                              & 0.04          & 0.01            & 0.12\\   
                                                & (0.07)  & (0.09)                            & (0.08)        & (0.07)          & (0.11)\\   
      Number of Counties                        & 56      & 56                                & 54            & 56              & 56\\  
      \midrule
      \emph{Fixed-effects}\\
      County                                    & Yes     & Yes                               & Yes           & Yes             & Yes\\  
      Year                                      & Yes     & Yes                               & Yes           & Yes             & Yes\\  
      \midrule
      \emph{Fit statistics}\\
      Observations                              & 314     & 314                               & 420           & 447             & 455\\  
      Dependent variable mean                   & 0.91    & 0.40                              & 4.5           & 3.5             & 3.6\\  
      F-test                                    & 27.4    & 10.5                              & 214.8         & 127.2           & 104.3\\  
      R$^2$                                     & 0.88    & 0.74                              & 0.98          & 0.97            & 0.96\\  
      \midrule \midrule
      \multicolumn{6}{l}{\emph{Clustered (County) standard-errors in parentheses}}\\
      \multicolumn{6}{l}{\emph{Signif. Codes: ***: 0.01, **: 0.05, *: 0.1}}\\
   \end{tabular}
   \begin{tablenotes}
   \item The sample consists of unbalanced panel data at the county-year from 2010 - 2022 for all observed counties within California. Effects are relative to two years prior to the adoption of charging station subsidies. Relative periods are binned at four years prior to the first change in charging station subsidy magnitude. Coefficients are weighted in proportion to county population. County-year observations including and after counties adopt charging station subsidies are excluded. Standard errors are clustered at the county level. Sample sizes vary due to missing values.
   \end{tablenotes}
   \end{threeparttable}
\end{table}

\begin{landscape}

\begin{table}[htbp]
   \caption{\label{tab:ols_sunab1} Effects of Charging Stations on Stations \\ and Sales Measures: Sun and Abraham Correction}
   \centering
   \begin{threeparttable}
   \begin{tabular}{lccccc}
      \tabularnewline \midrule \midrule
      Dependent Variables:    & \makecell[c]{Log\\(DC)}      & \makecell[c]{Log(Tesla Super \\ Charger Stations)} & \makecell[c]{Log\\(Sales)}    & \makecell[c]{Log\\(PHEV Sales)} & \makecell[c]{Log\\(BEV Sales)}\\  
      Model:                  & (1)          & (2)                               & (3)           & (4)             & (5)\\  
      \midrule
      \emph{Variables}\\
      DC Treated $\times$ Relative Year $=$ -3  to -12           & -0.03        & -0.13                             & -0.21$^{***}$ & -0.20$^{***}$   & -0.37$^{***}$\\   
                              & (0.10)       & (0.09)                            & (0.06)        & (0.07)          & (0.09)\\   
      DC Treated $\times$ Relative Year $=$ -2             & -0.004       & -0.02                             & -0.02         & 0.002           & -0.03\\   
                              & (0.06)       & (0.06)                            & (0.04)        & (0.05)          & (0.06)\\   
      DC Treated $\times$ Relative Year $=$ 0              & 0.07         & -0.03                             & -0.03         & 0.08            & -0.07\\   
                              & (0.08)       & (0.09)                            & (0.05)        & (0.06)          & (0.08)\\   
      DC Treated $\times$ Relative Year $=$ 1              & -0.010       & -0.15                             & -0.02         & 0.17$^{***}$    & -0.04\\   
                              & (0.09)       & (0.10)                            & (0.08)        & (0.06)          & (0.12)\\   
      DC Treated $\times$ Relative Year $=$ 2              & 0.36$^{**}$  & 0.01                              & 0.03          & 0.24$^{**}$     & -0.02\\   
                              & (0.15)       & (0.18)                            & (0.10)        & (0.11)          & (0.14)\\   
      DC Treated $\times$ Relative Year $=$ 3 to 12              & 0.50$^{***}$ & 0.02                              & 0.04          & 0.24$^{**}$     & 0.006\\   
                              & (0.17)       & (0.23)                            & (0.11)        & (0.12)          & (0.16)\\   
      Number of Counties      & 56           & 56                                & 54            & 56              & 56\\  
      \midrule
      \emph{Fixed-effects}\\
      County                  & Yes          & Yes                               & Yes           & Yes             & Yes\\  
      Year                    & Yes          & Yes                               & Yes           & Yes             & Yes\\  
      \midrule
      \emph{Fit statistics}\\
      Observations            & 479          & 479                               & 554           & 581             & 589\\  
      Dependent variable mean & 1.5          & 0.74                              & 5.2           & 4.2             & 4.4\\  
      F-test                  & 21.0         & 6.0                               & 114.1         & 64.0            & 60.2\\  
      R$^2$                   & 0.95         & 0.83                              & 0.99          & 0.98            & 0.98\\  
      \midrule \midrule
      \multicolumn{6}{l}{\emph{Clustered (County) standard-errors in parentheses}}\\
      \multicolumn{6}{l}{\emph{Signif. Codes: ***: 0.01, **: 0.05, *: 0.1}}\\
   \end{tabular}
   \begin{tablenotes}
   \item The sample consists of unbalanced panel data at the county-year from 2010 - 2022 for all observed counties within California. Effects are relative to one year prior to the adoption of charging station subsidies, defined as charging station subsidy magnitude changing from zero to positive. Relative periods are binned at two years prior to the first change in charging station subsidy magnitude and three years after. County-year observations among counties that always have a positive charging station subsidy magnitude when observed are dropped. Standard errors are clustered at the county level. All county-year observations are included for counties that do not adopt charging station subsidies during the sample period. Sample sizes vary due to missing values.
   \end{tablenotes}
   \end{threeparttable}
\end{table}


\end{landscape}



\clearpage

\appendix
\section*{Appendix}
\section{Level 2 Baseline Results}

Table A1 reports OLS estimation results for the same six specifications as in Table 4 where the independent variable is DC subsidies, the Column 1 variable is Level 2 stations, and the Column 2 variable is Tesla Destination stations (classified as Level 2). The baseline coefficient in Table A1 Column 1 (0.025) is 0.01 less than that in Table 4 (0.026). While the baseline + controls specification in Table A1 has a greater Column 1 coefficient than the baseline + controls specification in Table 4, the difference is also fairly minimal, suggesting that the control variables do not have a significant influence on Level 2 charging stations. Similar to Table 4, the Column 1 coefficients of Rows 3-6 in Table A1 — 0.022, 0.016, 0.021, 0.021 — respectively, are smaller and less statistically significant than the coefficient of 0.028 on the baseline + controls specification. Similar to Table 4, the only specification lacking statistical significance — the disadvantaged specification — also has the least number of observations (249). This finding provides further suggestive evidence that a lack of sufficient observations is one potential explanation for the smaller coefficients on Rows 3-6 compared to the baseline + controls specification. While my results in this regard do not align with theoretical expectations, future research should utilize more observations to examine how the effect of charging station subsidies on charging station supply varies over heterogeneous treatment groups, either from an extended sample period or by including data beyond California. 

Similar to Table 4, Table A1 has positive and significant coefficients in Columns 3 and 4 on the baseline and control specifications. However, as in Table 4, I do not assign causal interpretations due to heterogeneity. 

\section{Level 2 Robustness Checks}
\subsection{Event Study}
Using the same event study specification as Table 5, Table A2 reports coefficients on the interaction term between treatment status and relative years in the pre-treatment period and the relative post-treatment year $\tau = -1$. A county is defined to be treated if it adopts Level 2 subsidies in the period 2010 - 2022 and untreated if it does not. The omitted reference year is $\tau = -2$. As shown in Columns 1 and 2, I document null pre trends in Level 2 stations and Tesla Destination Stations (classified as DC stations). The lack of anticipatory effects supports the parallel trends assumption in the staggered difference-in-differences specification (see Equation 1). 

Similar to Table 5, I document anticipatory effects on Columns 3 and 5 and growing EV sales trends. Specifically, in Log(Sales) the binned value coefficient of -0.17 (std. err. 0.09) increases by 0.13 to -0.04 (std. Err. 0.04); and in Log(BEV Sales) the binned value coefficient of -0.40 (std. err. 0.15) increases by 0.27 to -0.13 (std. err. 0.07). While the lack of significant pre-trends in Column 4 (PHEV sales) deviates from the pattern of pre-trends in aggregate EV sales and BEV sales, it may be a result of treatment counties having a disproportionately higher number of BEVs than PHEVs (Zhou, n.d.). Since on average BEVs are more expensive than PHEVs, this implies that treatment counties tend to be relatively wealthy, which is consistent with the observation that treatment counties have a greater EV presence and capacity to afford EVs. 

\subsection{Sun and Abraham Correction for Staggered Treatment Timing}
Table A3 reports results from the same Sun and Abraham correction as in Table 6, but with Level 2 stations and Tesla Destination Stations. In contrast to Table 6, I find significant pre trends for the binned value from $\tau = -3$ to $\tau = -12$ in Tesla Destination Stations: 0.22 (std. err. 0.09). Similar to Table 6, I find that the coefficient on Level 2 stations decreases prior to the treatment by 0.08 from 0.06 (std. err. 0.08) to -0.02 (std. err. 0.05), though there are no significant observed pre-trends. The coefficient of 0.22 (std. err. 0.11) 2 years and 0.34 (std. err. 0.13) 3 years following subsidy adoption demonstrates that subsidy adoption increases Level 2 station supply by 24.6\% (40.5\%) two (three) years following subsidy adoption. Consistent with prior results, this finding shows that the impact of the subsidy on stations is not immediate, and that it increases over time. 

Also similar to Table 6, I document anticipatory effects on Columns 3, 4, and 5 and positive EV sales trends. Specifically, in Log(Sales) the binned value coefficient of -0.20 (std. err. 0.06) increases by 0.13 to -0.07 (std. err. 0.04); in Log(PHEV Sales) the binned value coefficient of -0.25 (std. err. 0.06) increases by 0.20 to -0.05 (std. err. 0.05); and in Log(BEV Sales) the binned value coefficient of -0.27 (std. err. 0.08) increases by 0.19 to -0.08 (std. err. 0.05). Still, there are no significant post-treatment effects of the subsidy on any sales measure.

\begin{landscape}

\setcounter{table}{0}
\renewcommand{\thetable}{A\arabic{table}}

\begin{table}

\caption{\label{tab:did2}Effects of Level 2 Subsidies on Level 2 Stations \\ and Sales Measures: Staggered Difference-in-Differences}
\centering
\begin{threeparttable}
\begin{tabular}[t]{lllll}
\toprule
  & Log(Lvl2) & Log(Tesla Destination Stations) & Log(PHEV Sales) & Log(BEV Sales)\\
\midrule
Baseline & 0.025 (0.012) ** & -0.009 (0.019) & 0.018 (0.009) ** & 0.028 (0.009) ***\\
Baseline + Controls & 0.028 (0.009) *** & 0.003 (0.020) & 0.022 (0.009) ** & 0.024 (0.009) **\\
High Pop & 0.022 (0.010) ** & 0.009 (0.025) & 0.010 (0.009) & 0.009 (0.007)\\
Disadvantaged & 0.016 (0.012) & 0.014 (0.029) & 0.011 (0.009) & -0.003 (0.008)\\
Urban & 0.021 (0.011) * & 0.002 (0.025) & 0.008 (0.008) & 0.010 (0.007)\\
\addlinespace
High GDP & 0.021 (0.011) * & 0.003 (0.026) & 0.010 (0.009) & 0.011 (0.007)\\
\bottomrule
\end{tabular}
\begin{tablenotes}
\item \textit{Note: } 
\item The sample restricts to an unbalanced panel of county-year observations for all observed counties in California. The time period covered is 2010 - 2022. The table presents the coefficients and standard errors on the treated x post treatment period interaction term. The treatment is defined as a county’s adoption of charging station subsidies. Treated counties adopt charging station subsidies within the time period 2010-2022 and the post treatment period is the period of time after the adoption of charging stations subsidies to 2022. Standard errors are clustered at the county level. Charging station counts and sales measures are log transformed. To account for possible zero values, one is added to variables prior to log transformation. Row 2 adds controls for GDP and purchase price subsidies. Row 3 subsets to counties whose population is greater than the median population among all California counties. Row 4 only considers counties classified as disadvantaged in 2020 in accordance with SB 535. Row 5 subsets to counties classified as rural in 2020 in accordance with the guidelines from the California State Association of Counties. Row 6 subsets to counties with GDP higher than the median GDP across all California counties in 2020. Tesla destination stations are classified as level 2 stations and are a part of Tesla's proprietary charging network. Tesla superchargers are classified as DC stations and are a part of Tesla's proprietary charging network.
\end{tablenotes}
\end{threeparttable}
\end{table} 

\clearpage

\begin{table}[htbp]
   \caption{\label{tab:ols2} Effects of Level 2 Charging Stations on Level 2 Stations \\ and Sales Measures: OLS Results}
   \centering
   \begin{threeparttable}
   \begin{tabular}{lccccc}
      \tabularnewline \midrule \midrule
      Dependent Variables:                           & \makecell[c]{Log\\(Level 2)} & \makecell[c]{Log(Tesla \\ Destination Stations)} & \makecell[c]{Log\\(Sales)}  & \makecell[c]{Log\\(PHEV Sales)} & \makecell[c]{Log\\(BEV Sales)}\\  
      Model:                                         & (1)          & (2)                             & (3)         & (4)             & (5)\\  
      \midrule
      \emph{Variables}\\
      Level 2 Treated $\times$ Relative Year $=$ -4 to -12 & -0.09        & 0.36                            & -0.17$^{*}$ & -0.10           & -0.40$^{**}$\\   
                                                     & (0.10)       & (0.28)                          & (0.09)      & (0.09)          & (0.15)\\   
      Level 2 Treated $\times$ Relative Year $=$ -3  & -0.03        & 0.0009                          & -0.04       & 0.006           & -0.13$^{*}$\\   
                                                     & (0.06)       & (0.09)                          & (0.04)      & (0.06)          & (0.07)\\   
      Level 2 Treated $\times$ Relative Year $=$ -1  & 0.08         & 0.01                            & 0.14$^{**}$ & 0.12            & 0.20$^{***}$\\   
                                                     & (0.05)       & (0.15)                          & (0.06)      & (0.08)          & (0.07)\\   
      Number of Counties                             & 56           & 56                              & 54          & 56              & 56\\  
      \midrule
      \emph{Fixed-effects}\\
      County                                         & Yes          & Yes                             & Yes         & Yes             & Yes\\  
      Year                                           & Yes          & Yes                             & Yes         & Yes             & Yes\\  
      \midrule
      \emph{Fit statistics}\\
      Observations                                   & 314          & 314                             & 420         & 447             & 455\\  
      Dependent variable mean                        & 0.91         & 1.0                             & 4.5         & 3.5             & 3.6\\  
      F-test                                         & 27.6         & 19.9                            & 216.2       & 128.0           & 106.1\\  
      R$^2$                                          & 0.88         & 0.84                            & 0.98        & 0.97            & 0.96\\  
      \midrule \midrule
      \multicolumn{6}{l}{\emph{Clustered (County) standard-errors in parentheses}}\\
      \multicolumn{6}{l}{\emph{Signif. Codes: ***: 0.01, **: 0.05, *: 0.1}}\\
   \end{tabular}
   \begin{tablenotes}
   \item The sample consists of unbalanced panel data at the county-year from 2010 - 2022 for all observed counties within California. Effects are relative to two years prior to the adoption of charging station subsidies. Relative periods are binned at four years prior to the first change in charging station subsidy magnitude. Coefficients are weighted in proportion to county population. County-year observations including and after counties adopt charging station subsidies are excluded. Standard errors are clustered at the county level. Sample sizes vary due to missing values.
   \end{tablenotes}
   \end{threeparttable}
\end{table}
\clearpage

\begin{table}[htbp]
   \caption{\label{tab:ols_sunab2} Effects of Level 2 Charging Stations on Level 2 Stations \\ and Sales Measures: Sun and Abraham Correction}
   \centering
   \begin{threeparttable}
   \begin{tabular}{lccccc}
      \tabularnewline \midrule \midrule
      Dependent Variables:    & \makecell[c]{Log\\(Level 2)} & \makecell[c]{Log(Tesla \\ Destination Stations)} & \makecell[c]{Log\\(Sales)}    & \makecell[c]{Log\\(PHEV Sales)} & \makecell[c]{Log\\(BEV Sales)}\\  
      Model:                  & (1)          & (2)                             & (3)           & (4)             & (5)\\  
      \midrule
      \emph{Variables}\\
      Level 2 Treated $\times$ Relative Year $=$ -3 to -12             & 0.06         & 0.22$^{**}$                     & -0.20$^{***}$ & -0.25$^{***}$   & -0.27$^{***}$\\   
                              & (0.08)       & (0.09)                          & (0.06)        & (0.06)          & (0.08)\\   
      Level 2 Treated $\times$ Relative Year $=$ -2             & -0.02        & -0.01                           & -0.07$^{*}$   & -0.05           & -0.08\\   
                              & (0.05)       & (0.05)                          & (0.04)        & (0.05)          & (0.05)\\   
      Level 2 Treated $\times$ Relative Year $=$ 0              & 0.09         & 0.10$^{*}$                      & -0.11$^{**}$  & -0.12$^{**}$    & -0.06\\   
                              & (0.07)       & (0.05)                          & (0.05)        & (0.05)          & (0.07)\\   
      Level 2 Treated $\times$ Relative Year $=$ 1              & 0.01         & 0.08                            & -0.12         & -0.08           & -0.08\\   
                              & (0.08)       & (0.06)                          & (0.07)        & (0.05)          & (0.11)\\   
      Level 2 Treated $\times$ Relative Year $=$ 2              & 0.22$^{*}$   & 0.16                            & -0.11         & -0.07           & -0.02\\   
                              & (0.11)       & (0.10)                          & (0.10)        & (0.10)          & (0.12)\\   
      Level 2 Treated $\times$ Relative Year $=$ 3 to 12             & 0.34$^{**}$  & 0.25$^{**}$                     & -0.10         & -0.11           & 0.04\\   
                              & (0.13)       & (0.12)                          & (0.10)        & (0.10)          & (0.14)\\   
      Number of Counties      & 56           & 56                              & 54            & 56              & 56\\  
      \midrule
      \emph{Fixed-effects}\\
      County                  & Yes          & Yes                             & Yes           & Yes             & Yes\\  
      Year                    & Yes          & Yes                             & Yes           & Yes             & Yes\\  
      \midrule
      \emph{Fit statistics}\\
      Observations            & 479          & 479                             & 554           & 581             & 589\\  
      Dependent variable mean & 1.5          & 1.4                             & 5.2           & 4.2             & 4.4\\  
      F-test                  & 14.3         & 7.2                             & 75.6          & 45.7            & 38.6\\  
      R$^2$                   & 0.95         & 0.90                            & 0.99          & 0.98            & 0.98\\  
      \midrule \midrule
      \multicolumn{6}{l}{\emph{Clustered (County) standard-errors in parentheses}}\\
      \multicolumn{6}{l}{\emph{Signif. Codes: ***: 0.01, **: 0.05, *: 0.1}}\\
   \end{tabular}
   \begin{tablenotes}
   \item The sample consists of unbalanced panel data at the county-year from 2010 - 2022 for all observed counties within California. Effects are relative to one year prior to the adoption of charging station subsidies, defined as charging station subsidy magnitude changing from zero to positive. Relative periods are binned at two years prior to the first change in charging station subsidy magnitude and three years after. County-year observations among counties that always have a positive charging station subsidy magnitude when observed are dropped. Standard errors are clustered at the county level. All county-year observations are included for counties that do not adopt charging station subsidies during the sample period. Sample sizes vary due to missing values.
   \end{tablenotes}
   \end{threeparttable}
\end{table}


\end{landscape}

\begin{table}[htp]
 \centering
   \centering
    \caption{Variable Definitions: Data Dictionary}
    

\section*{Supplementary material}

**Table 4:** Variable Definitions: Data Dictionary

This table defines key variables used in the analysis.

| Variable name | Definition |
| --- | --- |
| County | County in California (source: CSAC) |
| Year | Year of observation |
| Log(Tesla Super Chargers)$_{ct}$ | Total number of Tesla Superchargers in a given county in a given year (source: NREL) |
| Log (Tesla Destination)$_{ct}$ | Total number of Tesla Destination Chargers in a given county in a given year (source: NREL) |
| Log(Lvl2)$_{ct}$ | Total number of Level 2 charging stations in a given county in a given year (source: NREL) |
| Log(DC)$_{ct}$ | Total number of DC fast charging stations in a given county in a given year (source: NREL) |
| Log(Total Stations)$_{ct}$ | Total stations (including Level 2 and DC) in a given county in a given year (source: NREL) |
| Log(BEV Sales)$_{ct}$ | Total battery electric vehicle (BEV) sales in a given county in a given year (source: CEC) |
| Log(PHEV Sales)$_{ct}$ | Total plug in hybrid (PHEV) sales in a given county in a given year (source: CEC) |
| Log(Sales)$_{ct}$ | Total electric vehicle sales in a given county in a given year (source: CEC) |
| BEV Sales Pct Change$_{ct}$ | Year over year percentage change in BEV sales in a given county in a given year (source: CEC) |
| PHEV Sales Pct Change$_{ct}$ | Year over year percentage change in PHEV sales in a given county in a given year (source: CEC) |
| Total Sales Pct Change$_{ct}$ | Year over year percentage change in EV sales in a given county in a given year (source: CEC) |
| BEV Sales Nominal Change$_{ct}$ | Year over year nominal change in BEV sales in a given county in a given year (source: CEC) |
| PHEV Sales Nominal Change$_{ct}$ | Year over year nominal change in PHEV sales in a given county in a given year (source: CEC) |
| Total Sales Nominal Change$_{ct}$ | Year over year nominal change in EV sales in a given county in a given year (source: CEC) |

| | |
|---|---|
| CA Sales$_t$ | Total EV sales in California in a given year (source: [CEC](#)) |
| Log(Purchase Price)$_{ct}$ | Total amount of purchase price subsidies available in a given county in a given year (source: author's calculations) |
| Log(Pop)$_{ct}$ | Control variable; population of a given county in a given year (source: [California Department of Finance](#)) |
| Log(GDP)$_{ct}$ | Control variable; gross domestic product of a given county in a given year. GDP is given in current dollars (source: [US Bureau of Economic Analysis](#)) |
| High GDP$_{ct}$ | Indicator variable that takes a value of one if a given county has GDP greater than the median GDP across all California counties that year. GDP is given in current dollars (source: author's calculations) |
| Disadvantaged$_{ct}$ | Indicator variable that takes a value of one if a given county is classified as "disadvantaged" in accordance with SB 535 and zero otherwise. In SB 535, CalEPA formally designates four categories of geographic areas as disadvantaged: i) Census tracts receiving the highest 25 percent of overall scores in CalEnviroScreen 4.0; ii) Census tracts lacking overall scores in CalEnviroScreen 4.0 due to data gaps, but receiving the highest 5 percent of CalEnviroScreen 4.0 cumulative pollution burden scores; iii) Census tracts identified in the 2017 DAC designation as disadvantaged, regardless of their scores in CalEnviroScreen 4.0; iv) Lands under the control of federally recognized tribes (source: [California OEHHA](#)) |
| Never Treated$_{ct}$ | Indicator variable that takes a value of one if no charging station subsidies of any type were available in a given county in any year from 2010 – 2022 and zero otherwise (source: author's calculations) |
| Lvl2 Sub Pct Chg$_{ct}$ | Year over year percentage change in the amount of Level 2 station subsidies available in a given county in a given year (source: author's calculations) |

| | |
|---|---|
| DC Sub Pct Chg$_{ct}$ | Year over year percentage change in the amount of DC station subsidies available in a given county (source: author's calculations) |
| Lvl2 Pct Chg$_{ct}$ | Year over year percentage change in the amount of Level 2 stations in a given county in a given year (source: author's calculations) |
| DC Pct Chg$_{ct}$ | Year over year percentage change in the amount of DC stations in a given county in a give year (source: author's calculations) |
| Median Pop$_t$ | Median population of all California counties in a given year (source: author's calculations) |
| High Population$_{ct}$ | Indicator variable that takes a value of one if the population of a given county in a given year is higher than the median population of all California counties in that year and zero otherwise (source: author's calculations) |
| Log(Total Charging Sub)$_{ct}$ | Sum of all EV charging station subsidies (Level 2, DC Fast) in a given county in a given year (source: author's calculations) |
| Log(Lvl2S)$_{ct}$ | Total amount of Level 2 charging station subsidies available in a given county in a given year (source: author's calculations) |
| Log(DCS)$_{ct}$ | Total amount of DC fast charging station subsidies available in a given county in a given year (source: author's calculations) |
| EV Share$_t$ | The share of all California vehicles that is comprised of the population of EVs "on the road." Vehicle population counts vehicles whose registration is either current or less than 35 days expired (source: [CEC](CEC)) |
| ICE Share$_t$ | The share of all California vehicles that is comprised of the population of internal combustion engines (ICEs) "on the road." Vehicle population counts vehicles whose registration is either current or less than 35 days expired (source: [CEC](CEC)) |
| Unemployment Rate$_{ct}$ | Rate of unemployment in a given county in a given year. The unemployment rate represents the number unemployed as a percent of the |

| | |
|---|---|
| | labor force (source: California Employment and Development Department) |
| CA Unemployment Rate$_t$ | California-wide unemployment rate in a given year. The unemployment rate represents the number unemployed as a percent of the labor force (source: Federal Reserve Economic Data) |
| High Unemployment$_{ct}$ | Indicator variable that takes a value of one if a county has a greater unemployment in a given year than the California-wide unemployment rate in a given year and zero otherwise (source: California Employment and Development Department and Federal Reserve Economic Data) |
| Rural$_c$ | Indicator variable that takes a value of one if a county is defined as rural according to the California State Association of Counties (source: California State Association of Counties) |
| White$_c$ | Indicator variable that takes a value of one if the proportion of white individuals as a share of the total population in a given county is greater than that of California in a given year and zero otherwise (source: California Department of Finance) |



\end{document}